\documentclass[twocolumn,preprintnumbers,amssymb,amsmath,superscriptaddress,letterpaper,nofootinbib,showpacs]{revtex4}[10pt]

\usepackage{graphicx}
\usepackage{dcolumn}
\usepackage{bm}
\usepackage{natbib}
\usepackage{epsfig}
\usepackage{units}

\newcommand{\be}{\begin{equation}}
\newcommand{\ee}{\end{equation}}
\newcommand{\bea}{\begin{eqnarray}}
\newcommand{\eea}{\end{eqnarray}}

\def\lsim{\mathrel{\raise.3ex\hbox{$<$\kern-.75em\lower1ex\hbox{$\sim$}}}}
\def\gsim{\mathrel{\raise.3ex\hbox{$>$\kern-.75em\lower1ex\hbox{$\sim$}}}}

\begin{document}

\hspace*{130mm}{\large \tt FERMILAB-PUB-13-090-A}
\vskip 0.2in

\title{Dark matter and pulsar origins of the rising cosmic ray positron fraction in light of new data from AMS} 

\author{Ilias Cholis}
\email{cholis@fnal.gov}
\affiliation{Fermi National Accelerator Laboratory, Center for Particle Astrophysics, Batavia, Illinois, 60510, USA}
\author{Dan Hooper}
\email{dhooper@fnal.gov}
\affiliation{Fermi National Accelerator Laboratory, Center for Particle Astrophysics, Batavia, Illinois, 60510, USA}
\affiliation{University of Chicago, Department of Astronomy and Astrophysics, Chicago, Illinois, 60637, USA}
\date{\today}

\begin{abstract}

The rise of the cosmic ray positron fraction with energy, as first observed with high confidence by \textit{PAMELA}, implies that a large flux of high energy positrons has been recently (or is being currently) injected into the local volume of the Milky Way. With the new and much more precise measurement of the positron fraction recently provided by \textit{AMS}, we revisit the question of the origin of these high energy positrons. We find that while some dark matter models (annihilating directly to electrons or muons) no longer appear to be capable of accommodating these data, other models in which $\sim$1-3 TeV dark matter particles annihilate to unstable intermediate states could still be responsible for the observed signal. Nearby pulsars also remain capable of explaining the observed positron fraction. Future measurements of the positron fraction by \textit{AMS} (using a larger data set), combined with their anticipated measurements of various cosmic ray secondary-to-primary ratios, may enable us to further discriminate between these remaining scenarios.

\end{abstract}


\maketitle

\section{Introduction}
\label{sec:intro}

Recently, the \textit{AMS} collaboration published its measurement of the cosmic ray positron fraction over the range of 0.5 to 350 GeV~\cite{AMS02}.
Their findings confirm with unprecedented precision earlier measurements from the \textit{PAMELA}~\cite{Picozza:2006nm} and \textit{Fermi}~\cite{Gehrels:1999ri} collaborations, which had each reported a clear rise in the positron fraction at energies above $\sim$10 GeV~\cite{Adriani:2008zr}~(hints of such a rise were also present in data from \textit{HEAT}~\cite{heat} and \textit{AMS-01}~\cite{AMS}). At the time, \textit{PAMELA}'s observation generated a great deal of interest and speculation as to the origin of the high energy positrons. Leading proposals put forth to explain this observation included dark matter (DM) particles annihilating or decaying in the Galactic Halo~\cite{Bergstrom:2008gr, Cirelli:2008jk, Cholis:2008hb, Barger:2008su, Cirelli:2008pk, Nelson:2008hj, ArkaniHamed:2008qn, Cholis:2008qq, Nomura:2008ru, Yin:2008bs, Harnik:2008uu, Fox:2008kb, Pospelov:2008jd, MarchRussell:2008tu,Chang:2011xn}, and nearby pulsars injecting high energy positrons into the interstellar medium~\cite{Hooper:2008kg, Yuksel:2008rf, Profumo:2008ms, Malyshev:2009tw, Grasso:2009ma}. An alternative explanation is that near-by supernova remnants could be accelerating electrons, positrons, produced from the decay of $\pi^{\pm}$ created in hadronic interactions of accelerated protons by the same source \cite{Blasi:2009hv, Mertsch:2009ph}.     

Despite providing valuable information, the measurements provided by \textit{PAMELA} and \textit{Fermi} were not sufficient to discriminate between DM and pulsar origins of the rising positron fraction. The much higher precision measurement of the positron fraction by \textit{AMS}, however, brings new and important information to bear on this question. In this article, we make use of this new data and revisit both annihilating DM's and pulsars' as potential sources of the observed high energy cosmic ray positrons. We find that DM particles which annihilate directly to $e^+ e^-$ or $\mu^+ \mu^-$ can no longer accommodate the observed positron fraction. However, DM particles with a mass of $\sim$1-3 TeV annihilating to intermediate states which then decay to muons or charged pions could potentially provide a good fit. Pulsars also continue to represent a potentially viable explanation for the observed positrons.

The remainder of this article is structured as follows. In Sec.~\ref{sec:DM}, we discuss whether annihilating DM can account for \textit{AMS}'s measurement of the positron fraction, considering a variety of DM models and models of cosmic ray propagation. In Sec.~\ref{sec:Pulsars}, we discuss whether pulsars can account for the observed data. In comparing these scenarios, we find that the existing data from \textit{AMS} cannot yet definitively discriminate between the DM's and pulsars' origins of the observed positrons, although the range of models capable of accommodating the data is now significantly more constrained. With future data from \textit{AMS}, providing not only measurements of the positron fraction but also of various cosmic ray secondary-to-primary ratios, we expect to be able to further narrow the range of models potentially responsible for the rising positron fraction. In Sec.~\ref{sec:Conclusions}, we summarize our results and briefly discuss our current understanding of the possible origins of the observed cosmic ray positron fraction.


\section{Annihilating Dark Matter}
\label{sec:DM}


If annihilating DM particles are to account for the observed rise in the cosmic ray positron fraction, they must be quite heavy, certainly no less than 350 GeV. Furthermore, models which can also accommodate the smoothly varying and consistently hard spectrum of cosmic ray electrons and positrons as measured by \textit{Fermi} \cite{Ackermann:2010ij, Abdo:2009zk} and \textit{HESS} \cite{Collaboration:2008aaa, Aharonian:2009ah} typically feature DM particles with masses on the order of $\simeq$ 1 TeV or higher~\cite{Cholis:2008wq, Chen:2008qs, Nardi:2008ix, ArkaniHamed:2008qn}. 

In many well motivated models, DM particles annihilate with a cross section on the order of $\langle \sigma v \rangle \sim 10^{-26}$ cm$^{3}$/s. If the dark matter's annihilation cross section (as evaluated in the early universe) is much larger than this value, the DM particles would have been overly depleted in the early universe, resulting in a thermal relic abundance that is much smaller than the measured cosmological density. In contrast, in order for annihilating DM to account for the observed positron fraction, the DM particles must annihilate to leptonic final states with a cross section of $\langle \sigma v \rangle \sim 10^{-24}-10^{-23}$ cm$^{3}$/s~\cite{Cholis:2008wq, Bergstrom:2009fa, Finkbeiner:2010sm}. Various proposals to accommodate this very high annihilation rate have been put forth, including annihilation cross sections which are enhanced at the low velocities found in the Galactic Halo (such as by Sommerfeld enhancements)~\cite{ArkaniHamed:2008qn,Fox:2008kb,Zurek:2008qg,Chen:2008dh,Lattanzi:2008qa,Hisano:2004ds}, DM particles which are produced largely through non-thermal processes~\cite{Moroi:1999zb}, or DM whose annihilation rate is highly boosted as a result of larger than expected inhomogeneities in their spatial distribution.

Further restricting any DM scenarios that might potentially account for the observed positrons is \textit{PAMELA}'s measurement of the cosmic ray antiproton-to-proton ratio~\cite{antiprotonpamela}, which is consistent with astrophysical expectations. This observation strongly constrains the rate at which DM particles can annihilate to gauge bosons or quarks~\cite{Cirelli:2008pk, Donato:2008jk, Evoli:2011id}. Furthermore, an annihilation rate to quarks, gauge bosons, or taus that is comparable to the leptonic rate required to produce the observed positron fraction would also lead to an unacceptably high flux of prompt $\gamma$-rays from the Galactic Center~\cite{Cirelli:2009dv, Abramowski:2011hc, Abazajian:2011ak} and from the Inner Galaxy~\cite{Ackermann:2012rg, MaryamEtAl} (constraints have also been derived from observations of dwarf spheroidal galaxies~\cite{2004PhRvD..69l3501E, Ackermann:2011wa, GeringerSameth:2011iw,Cholis:2012am} and the isotropic diffuse $\gamma$-ray background~\cite{Abdo:2010dk, Hutsi:2010ai, Calore:2011bt}). As a result of these antiproton and $\gamma$-ray constraints, we are forced to consider DM candidates which annihilate dominantly to electrons and muons, $\chi \chi \longrightarrow e^{+} e^{-}$, $\chi \chi \longrightarrow \mu^{+} \mu^{-}$, or to intermediate particles that later decay to combinations of $e^{+} e^{-}$, $\mu^{+} \mu^{-}$, and $\pi^+ \pi^-$. In considering this latter case, known as eXciting Dark Matter (XDM) \cite{Finkbeiner:2007kk} (see also Ref.~\cite{ArkaniHamed:2008qn}), we make use of the analytic spectra described in Ref.~\cite{Cholis:2008vb}. 

\begin{figure*}
\begin{centering}
\includegraphics[width=3.30in,angle=0]{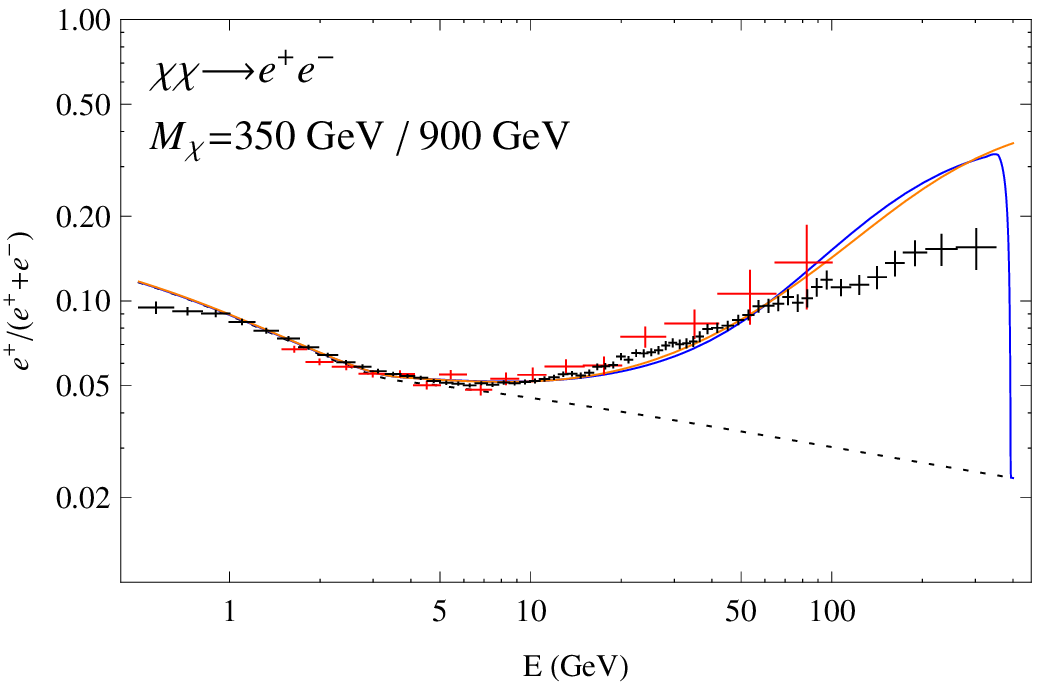}
\includegraphics[width=3.30in,angle=0]{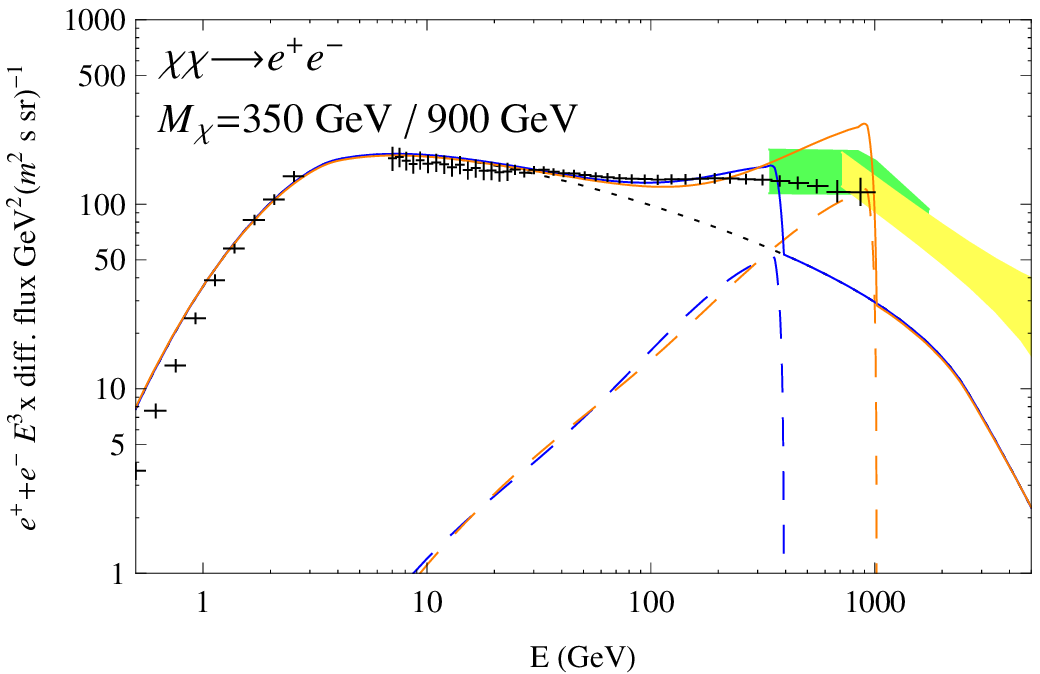} \\
\includegraphics[width=3.30in,angle=0]{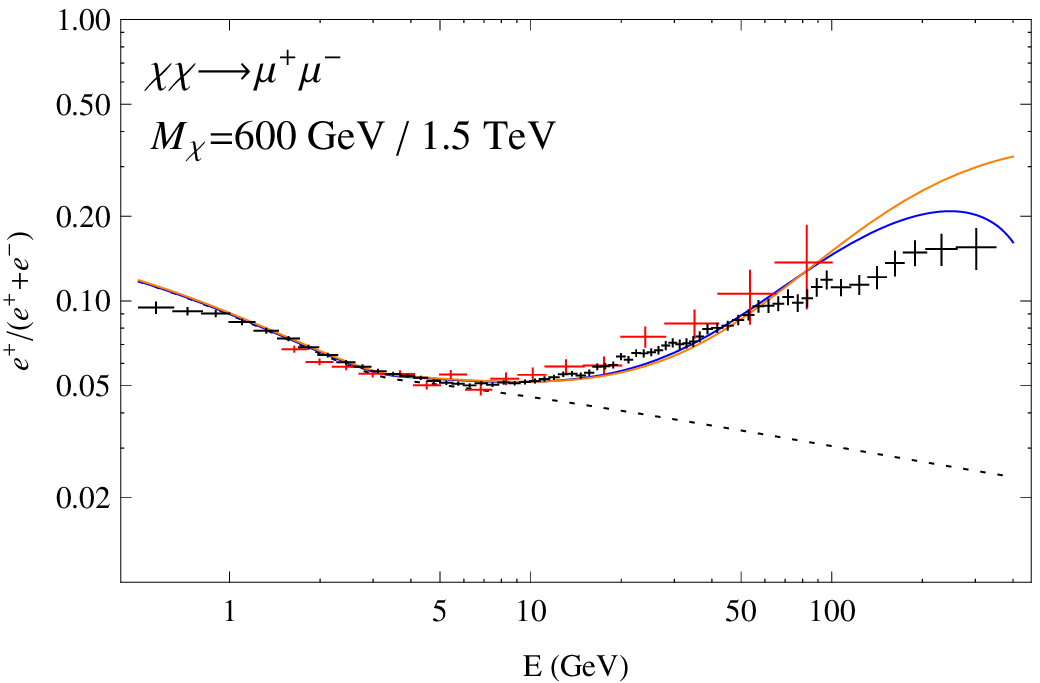}
\includegraphics[width=3.30in,angle=0]{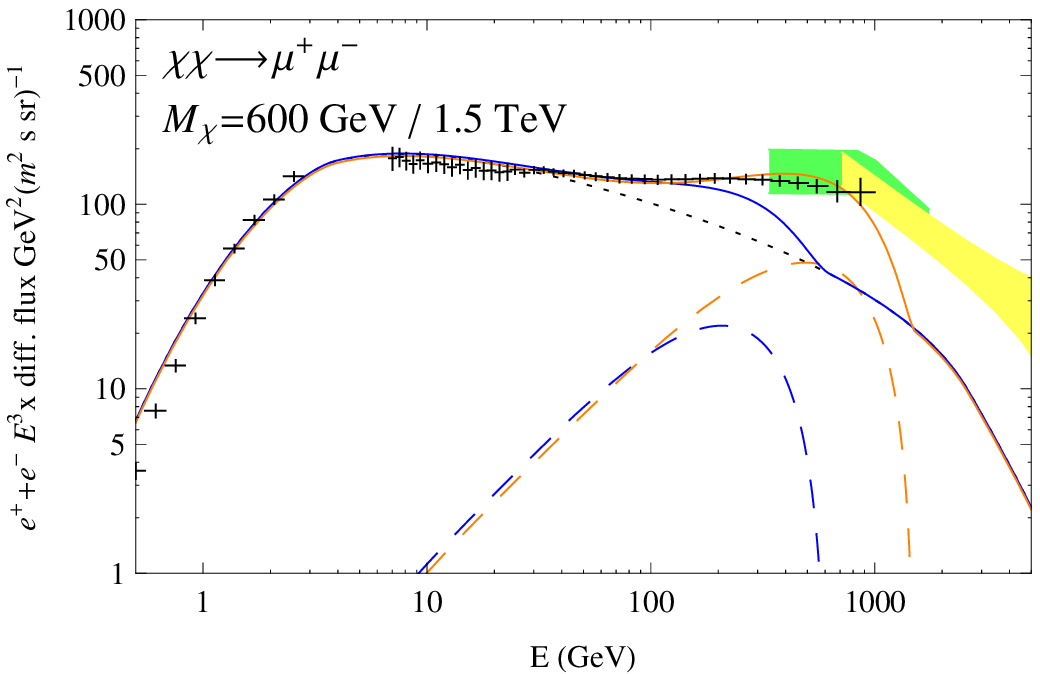} \\
\end{centering}
\caption{The predicted cosmic ray positron fraction (left) and electron+positron spectrum (right) in dark matter models annihilating to $e^+ e^-$ (top) and to $\mu^+ \mu^-$ (bottom). The error bars shown represent the positron fraction as measured by \textit{AMS} (black, left) and \textit{PAMELA} (red, left), and the electron+positron spectrum as measured by \textit{Fermi} and \textit{AMS-01} (black, right). In the \textit{Fermi} error-bars we do not include the overall shift from the energy resolution uncertainty. In each case, we have adopted a propagation model that provides a good fit to the various secondary-to-primary ratios as described in the text, and with a diffusion zone half-width of $L=4$ kpc. The expected backgrounds are shown as black dotted lines. For a given mass and channel we fit the annihilation cross-section to the \textit{AMS} positron fraction ratio data and also show the equivalent result for the total lepton flux. For each annihilation channel, we show results for two masses. For annihilations to $e^+ e^-$ and a mass of 350 GeV (900 GeV), we have used a thermally averaged annihilation cross section of $\langle\sigma v\rangle = 4.0\times 10^{-25}$ cm$^3$/s ($2.2 \times 10^{-24}$ cm$^3$/s). Our $\chi^{2}$/d.o.f. is 15.3(10.6) from the \textit{AMS} data and 6.5(9.2) from the \textit{Fermi} data. For annihilations to $\mu^+ \mu^-$ and a mass of 600 GeV (1.5 TeV), we have used a thermally averaged annihilation cross section of $1.6\times 10^{-24}$ cm$^3$/s ($8.5 \times 10^{-24}$ cm$^{3}$/s), resulting in a $\chi^{2}$/d.o.f. fit of 9.3(14.6) to the \textit{AMS} and 7.6(0.84) to the \textit{Fermi} data. These models can not provide a consistent picture to the combined AMS and Fermi lepton flux measurements.}
\label{fig:KRA4kpc_DM}
\end{figure*}

\begin{figure*}
\begin{centering}
\includegraphics[width=3.00in,angle=0]{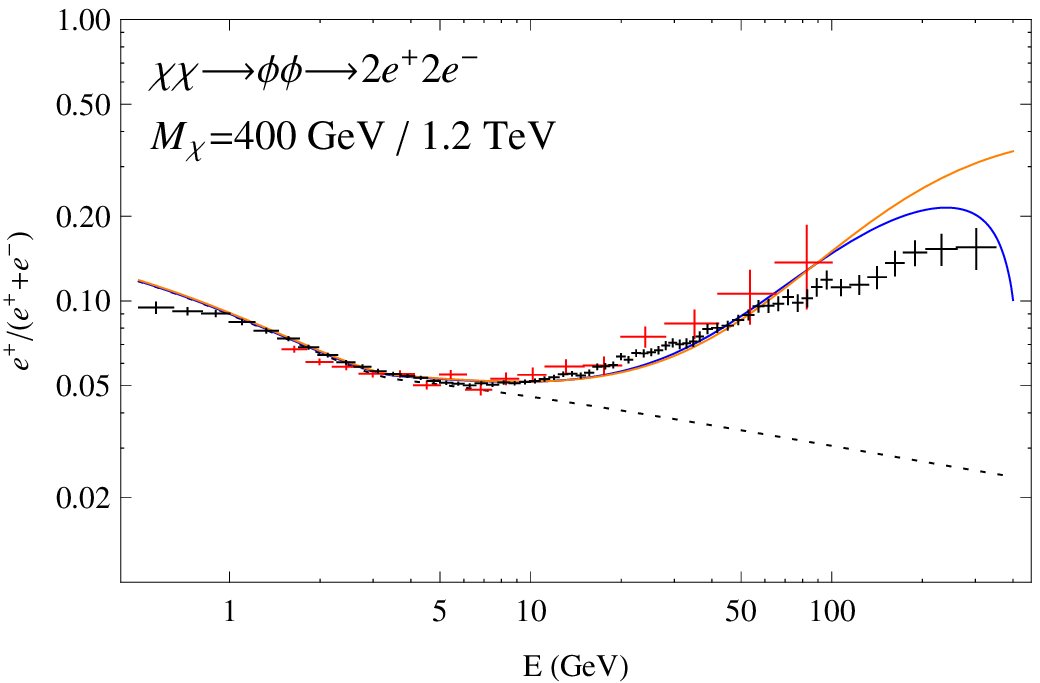}
\includegraphics[width=3.00in,angle=0]{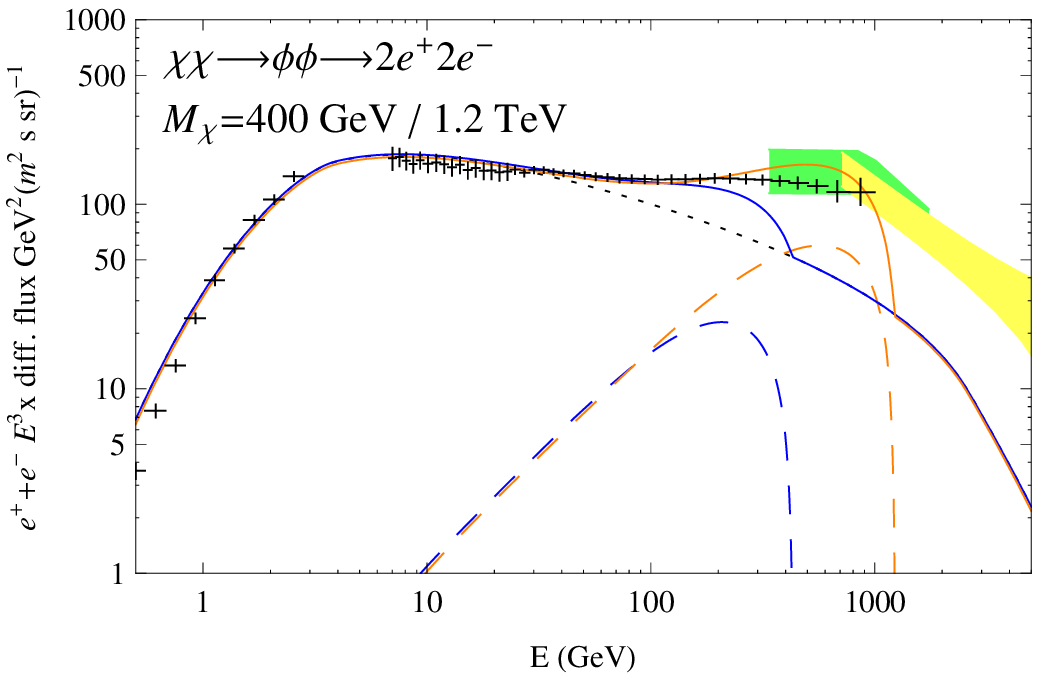} \\
\includegraphics[width=3.00in,angle=0]{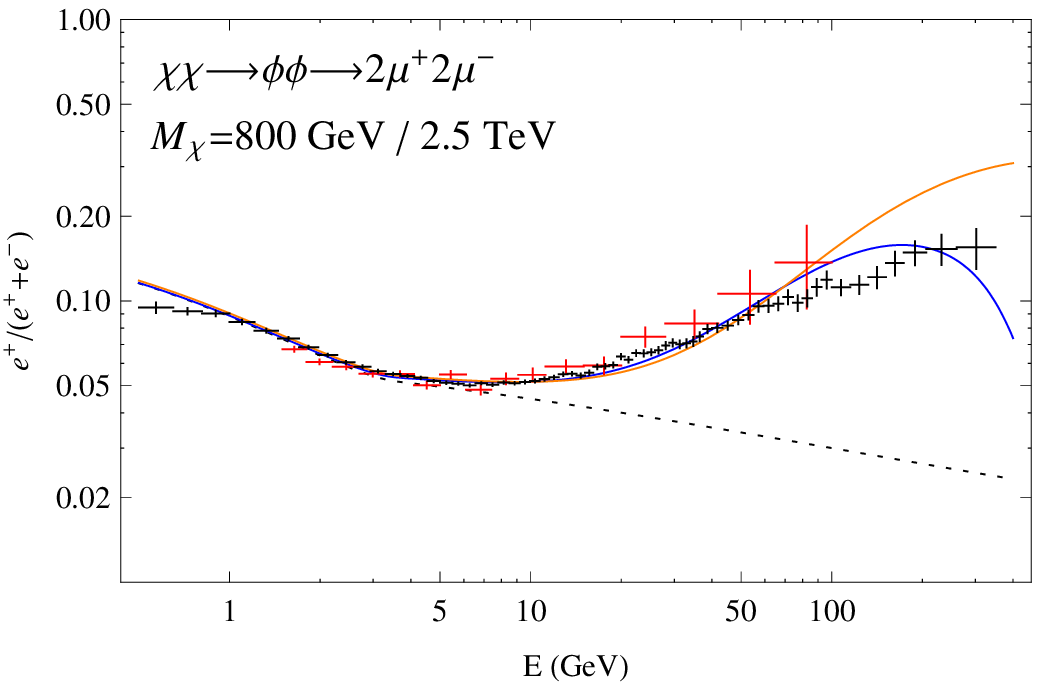}
\includegraphics[width=3.00in,angle=0]{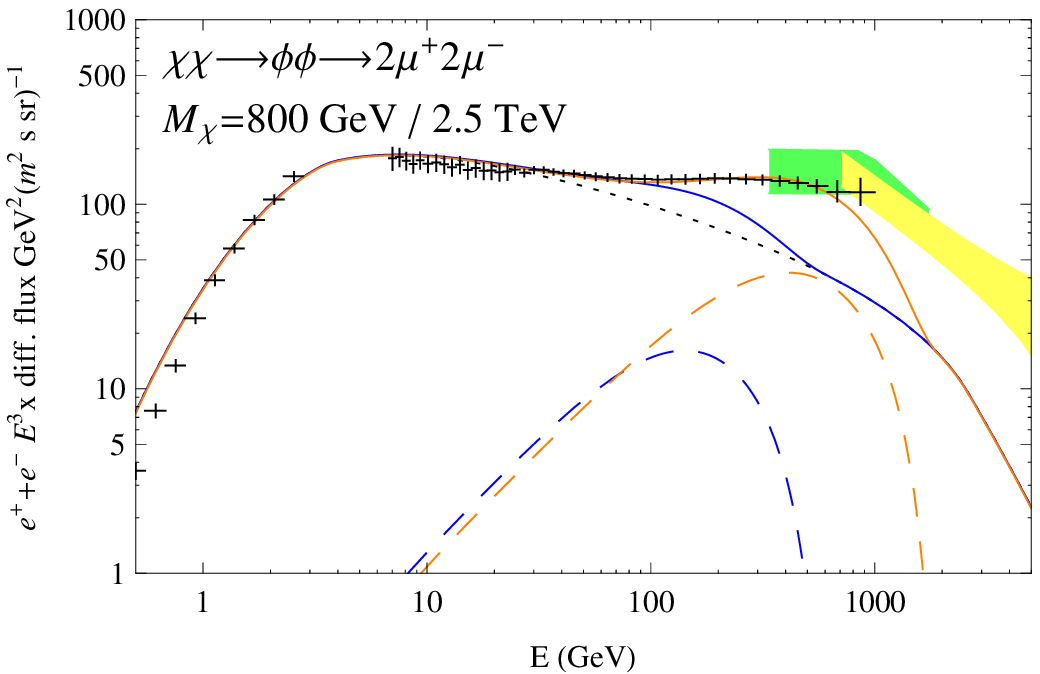} \\
\includegraphics[width=3.00in,angle=0]{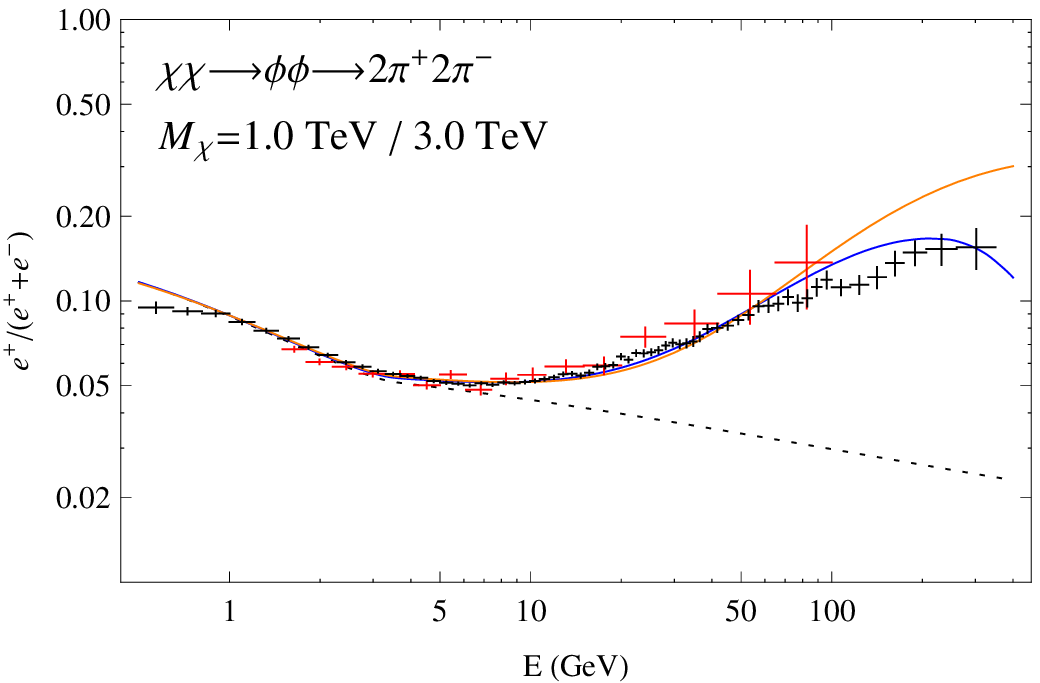}
\includegraphics[width=3.00in,angle=0]{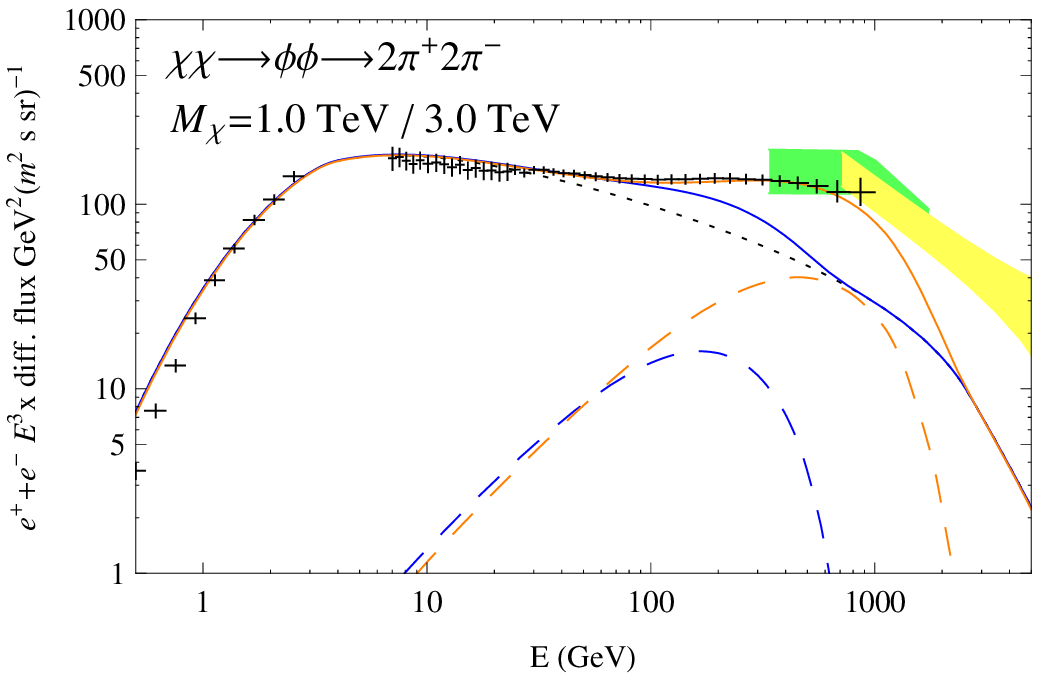} \\
\includegraphics[width=3.00in,angle=0]{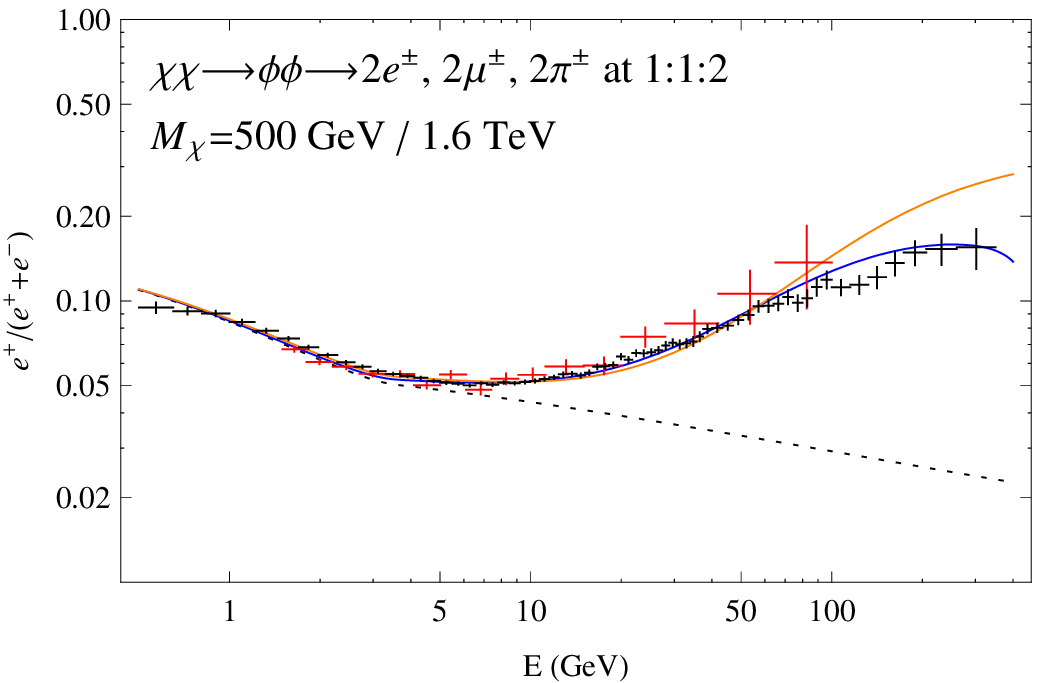}
\includegraphics[width=3.00in,angle=0]{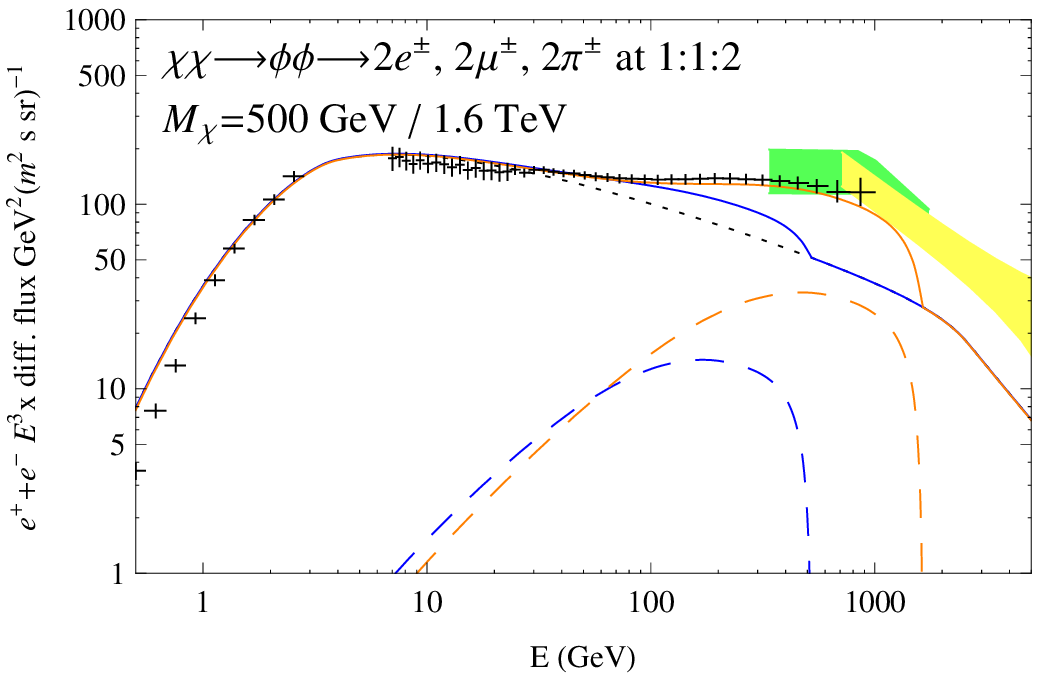}
\end{centering}
\caption{As in Fig.~1, but for dark matter which annihilates into a pair of intermediate states, $\phi$, which proceed to decay to $e^+ e^-$ (first row), to $\mu^+ \mu^-$ (second row), to $\pi^+ \pi^-$ (third row), and to a 1:1:2 ratio of $e^+ e^-$, $\mu^+ \mu^-$, and $\pi^+ \pi^-$ (fourth row). For annihilations to $2e^+ 2e^-$ and a mass of 400 GeV (1.2 TeV), we have used a thermally averaged annihilation cross section of $\langle \sigma v \rangle = 6.6\times 10^{-25}$ cm$^3$/s ($5.4 \times 10^{-24}$ cm$^3$/s); resulting in $\chi^{2}$/d.o.f. fit of 10.3(15.0) to the \textit{AMS} and 8.8(2.0) to the \textit{Fermi} data. For annihilations to $2\mu^+ 2\mu^-$ and a mass of 800 GeV (2.5 TeV), we have used a thermally averaged annihilation cross section of $3.4\times 10^{-24}$ cm$^3$/s ($2.6 \times 10^{-23}$ cm$^{3}$/s) with a $\chi^{2}$/d.o.f. of 5.1(14.0) and 12.6(0.64) to the \textit{AMS} and  \textit{Fermi} data respectively. For annihilations to $2\pi^+ 2\pi^-$ and a mass of 1.0 TeV (3.0 TeV), we have used a thermally averaged annihilation cross section of $5.8\times 10^{-24}$ cm$^3$/s ($4.1 \times 10^{-23}$ cm$^{3}$/s), giving $\chi^{2}$/d.o.f. fits of 3.7(11.7) to the \textit{AMS} and 19.4(0.61) to the \textit{Fermi} data. Finally for annihilations to a 1:1:2 ratio of $e^+ e^-$, $\mu^+ \mu^-$, and $\pi^+ \pi^-$ final states with a mass of 500 GeV (1.6 TeV), we have used a thermally averaged annihilation cross section of $1.7\times 10^{-24}$ cm$^3$/s ($1.2 \times 10^{-23}$ cm$^{3}$/s) which have a $\chi^{2}$/d.o.f. fits of 2.3(9.5) to the \textit{AMS} and 11.3(1.34) to the \textit{Fermi} data. While there  is a preference for DM models with softer annihilation $e^{\pm}$ spectra, for the given propagation/background assumptions all these models are excluded.}
\label{fig:KRA4kpc_DM_XDM}
\end{figure*}

High energy cosmic ray electrons and positrons undergo rapid energy losses as they diffuse through the interstellar medium. Above a few GeV, these energy losses are dominated by a combination of synchrotron and inverse Compton scattering. 
To model the propagation of electrons, positrons, and other cosmic rays, we use the GALPROP package v54 (see Refs.~\cite{Galprop1,website}, and references therein), which includes up-to-date information pertaining to the local interstellar radiation field and the distribution of gas in the Galaxy. 
The former is relevant for the calculation of $e^{\pm}$ energy losses via inverse Compton scattering, while the latter is relevant for the calculation of secondary cosmic ray species, produced in inelastic collisions of cosmic ray primaries with the interstellar gas. For synchrotron energy losses, we adopt a value of 5 $\mu$G for the local magnetic field. By fitting both stable and unstable secondary-to-primary ratios in the cosmic ray spectrum, we can constrain the distributions of the interstellar gas and cosmic ray sources, as well as the timescale that cosmic rays reside within the Galaxy~\cite{Strong:2007nh,Simet:2009ne, DiBernardo:2012zu}. Codes such as GALPROP and DRAGON~\cite{DRAGONweb} assume a simple diffusion zone with free escape boundary conditions (cosmic rays diffuse within the diffusion zone, but escape upon reaching any boundary of the zone). We take this zone to be a cylinder, extending a distance $L$ above and below the Galactic Plane, and radially 20 kpc from the Galactic Center. In this study, we vary the half-width of the diffusion zone over a range of values between $L=1$-8 kpc, in accordance with existing uncertainties~\cite{Simet:2009ne, Trotta:2010mx}. 
Values of $L<2$\,kpc are in tension with a combined analysis of cosmic and  $\gamma$-ray data \cite{Cholis:2011un}, while increasing $L$ well beyond $8$\,kpc
does not significantly alter our results.
With further information from \textit{AMS} (especially the measurement of the $^{10}$Be/$^{9}$Be ratio), this and other parameters of our propagation model will likely become significantly more constrained~\cite{Pato:2010ih}.

In Figs.~\ref{fig:KRA4kpc_DM} and~\ref{fig:KRA4kpc_DM_XDM}, for the choice of $L = 4$ kpc, we show the cosmic ray positron fraction $(\Phi_{e^{+}}/(\Phi_{e^{+}}+\Phi_{e^{-}}))$ and the total lepton spectrum ($E^{3}dN_{e^{\pm}}/dE_{e^{\pm}}$) predicted in a number of annihilating DM models. For each annihilation channel, we consider two values for the DM mass. The lower mass in each case is set to approximately the minimum value allowed by the new data from \textit{AMS} (in particular, by the lack of any strong spectral cut-off). The upper mass in each case was chosen with the cosmic ray electron spectrum (as measured by \textit{Fermi} and \textit{HESS}) in mind. As was found in previous fits to the \textit{PAMELA} data, we find that annihilation cross sections in the range of $\sim 4\times 10^{-25}-4\times 10^{-23}$ cm$^{3}$/s are required to accommodate the observed positron fraction. While changing the diffusion zone thickness or the radiation/magnetic field model can reduce this requirement to some extent, very large annihilation cross sections are a generic requirement of any annihilating dark matter model capable of generating the observed positron fraction. 

\begin{figure*}
\includegraphics[width=3.50in,angle=0]{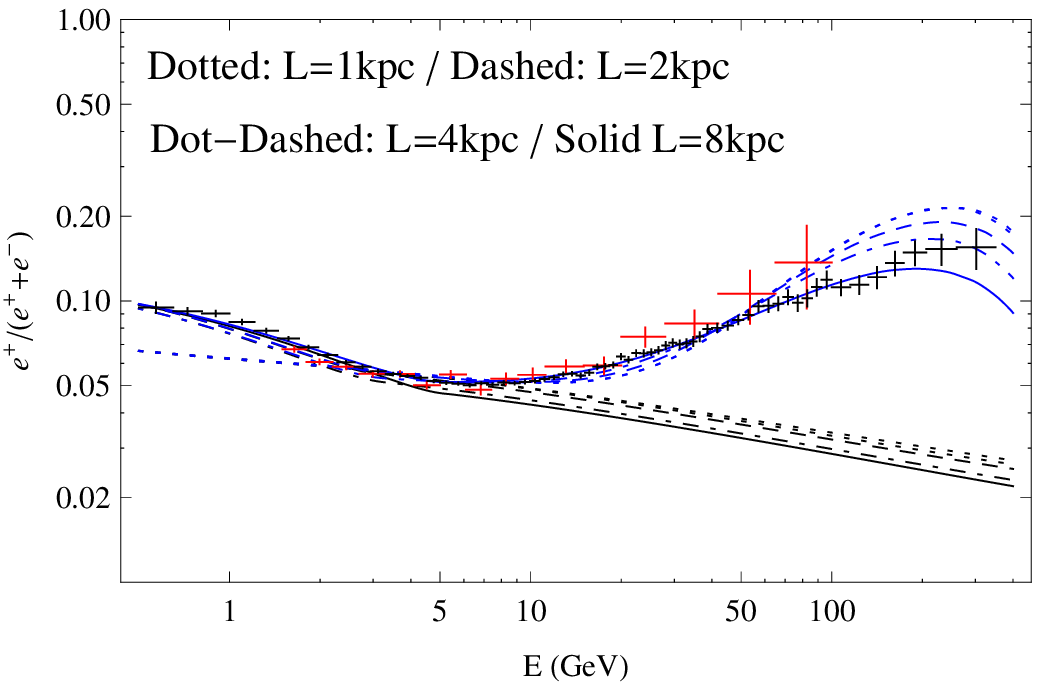}
\caption{The impact of the diffusion zone half-width, $L$, on the positron fraction in a representative DM model. In each case, we have chosen the diffusion coefficient to fit the non-leptonic background cosmic ray measurements. The dotted, dashed, dot-dashed, and solid lines correspond to $L=$1, 2, 4 and 8 kpc, respectively. The model used in this figure consists of a 1 TeV dark matter particle that annihilates to a pair of intermediate states which each decay to $\pi^{+}\pi^-$, with a cross section given by 16, 9.5, 5.7 and 3.5$\times 10^{-23}$cm$^{3}$/s and a $\chi^{2}$/d.o.f. of 16.6, 9.9, 3.7 and 1.18 for $L=$1, 2, 4 and 8 kpc, respectively.}
\label{fig:XDMmu_DiffZones}
\end{figure*}

\begin{figure*}
\begin{centering}
\includegraphics[width=3.00in,angle=0]{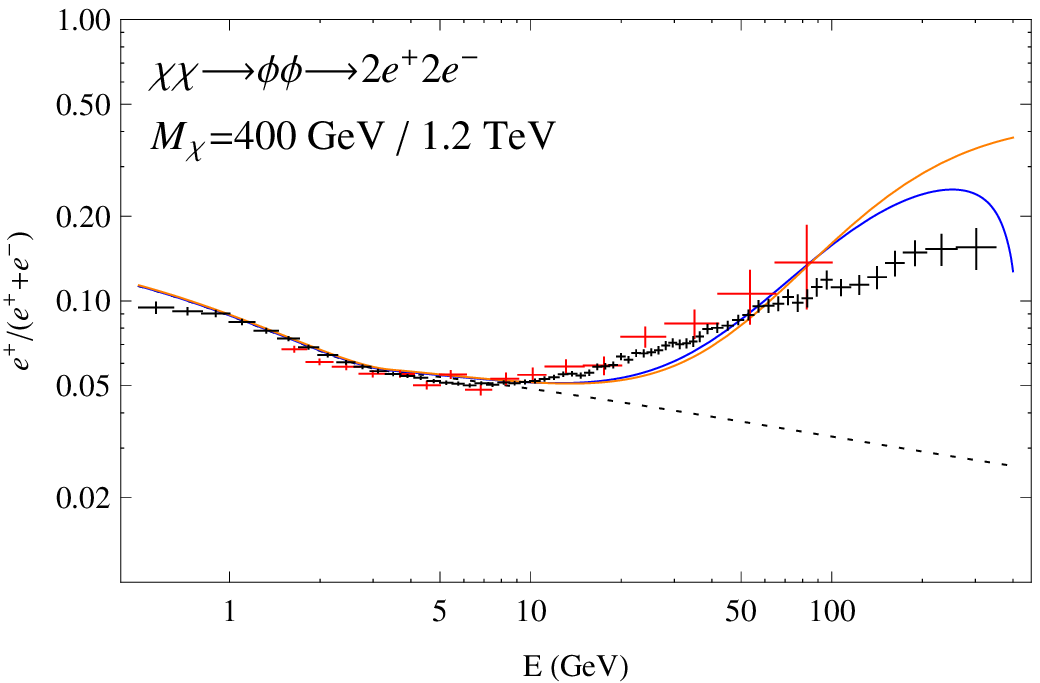}
\includegraphics[width=3.00in,angle=0]{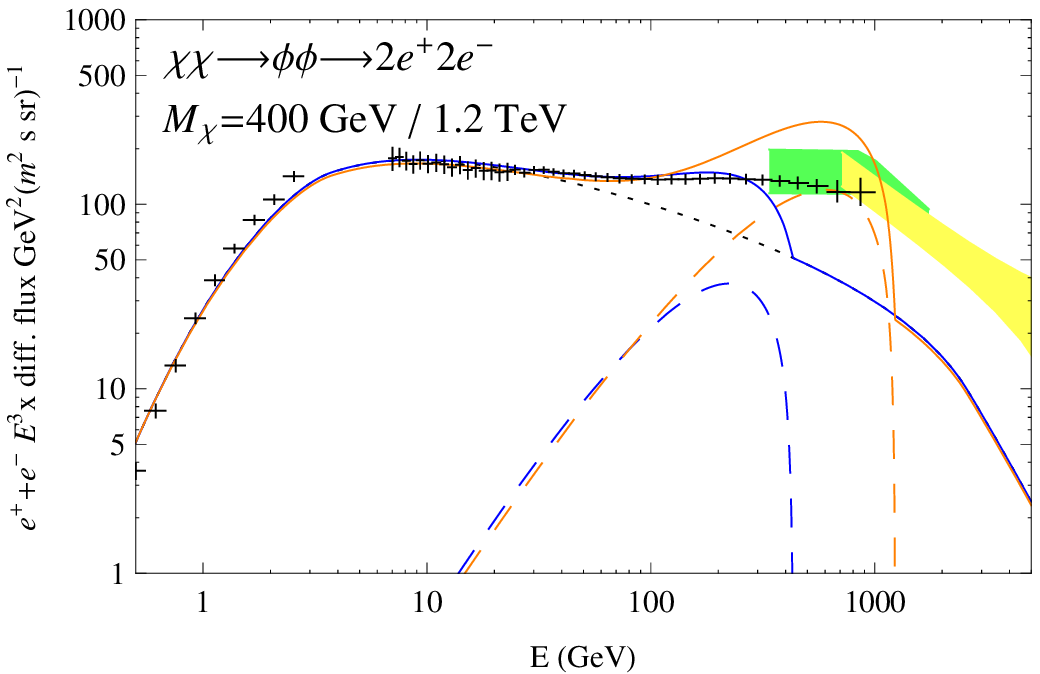} \\
\includegraphics[width=3.00in,angle=0]{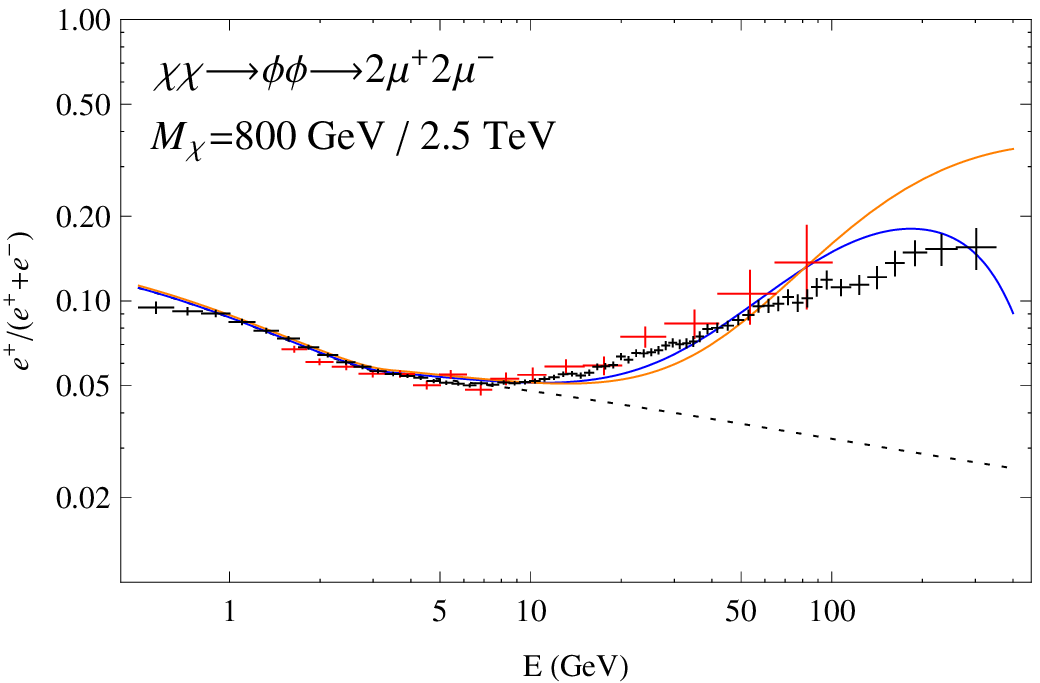}
\includegraphics[width=3.00in,angle=0]{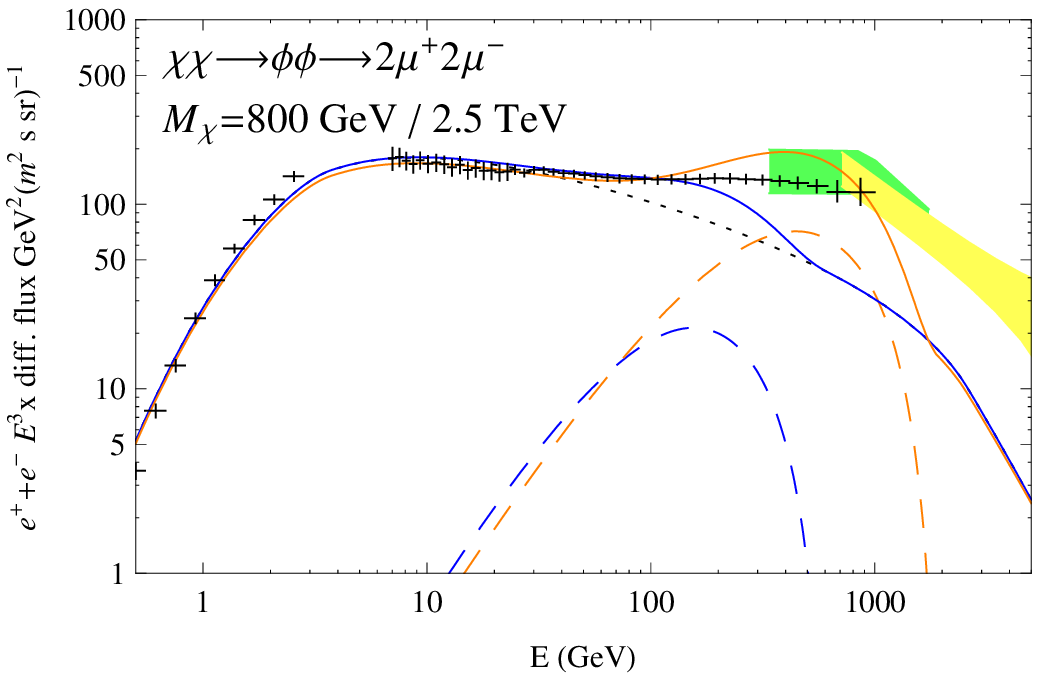} \\
\includegraphics[width=3.00in,angle=0]{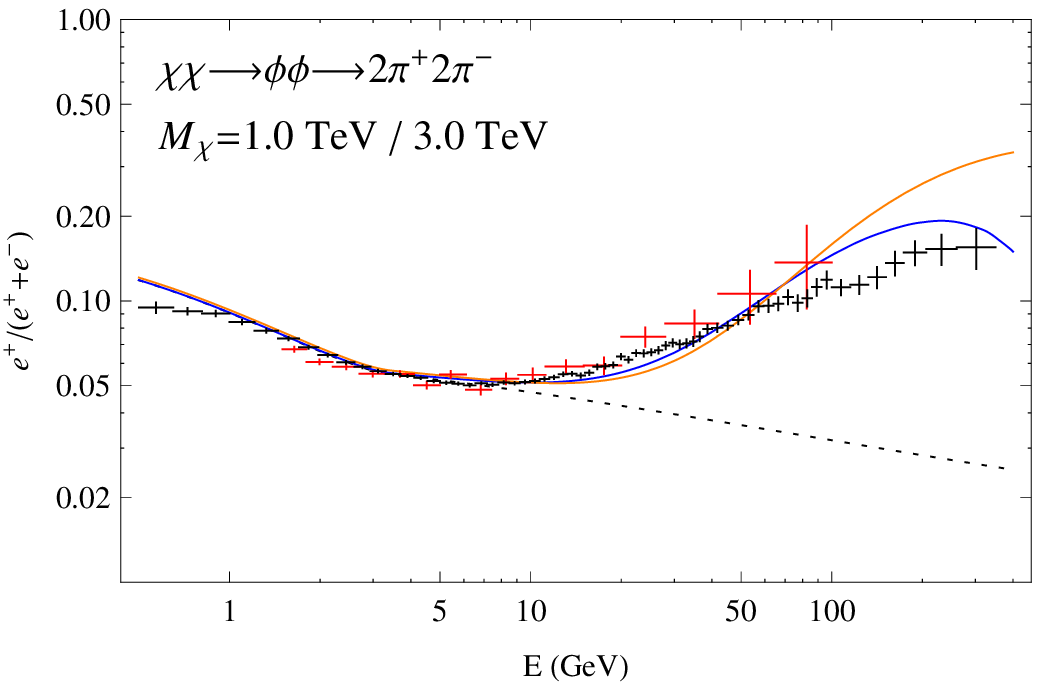}
\includegraphics[width=3.00in,angle=0]{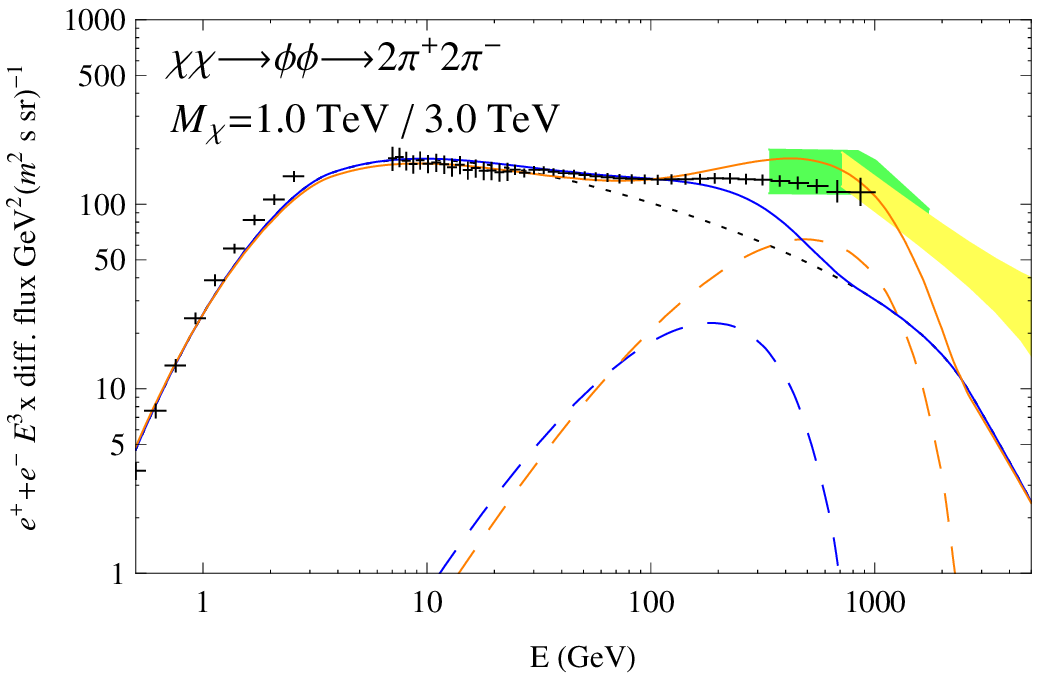} \\
\includegraphics[width=3.00in,angle=0]{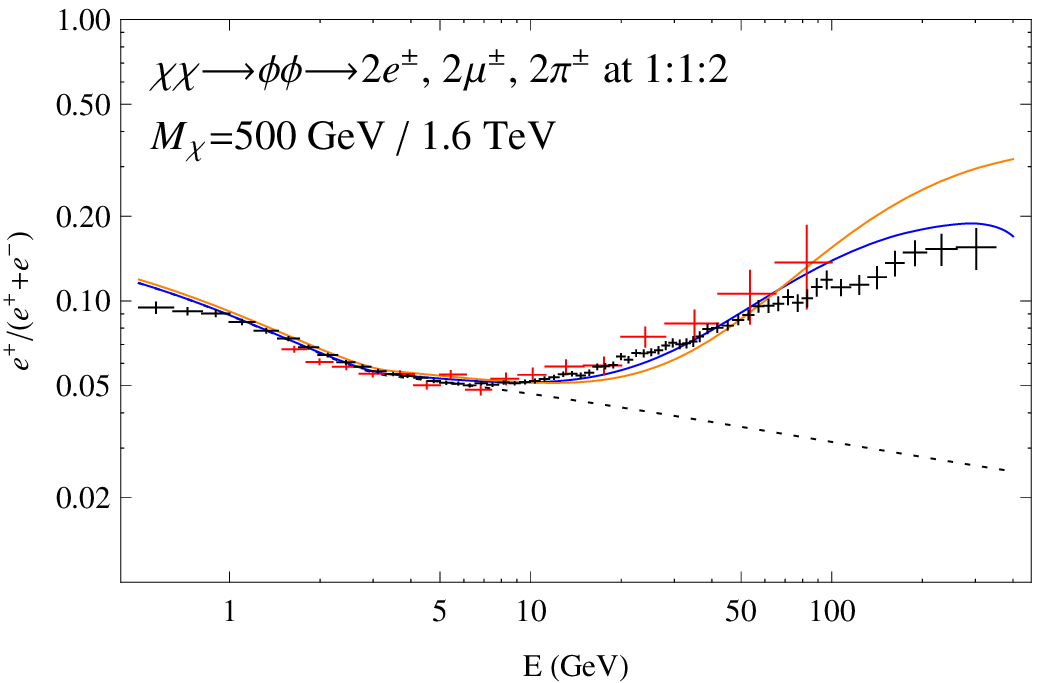}
\includegraphics[width=3.00in,angle=0]{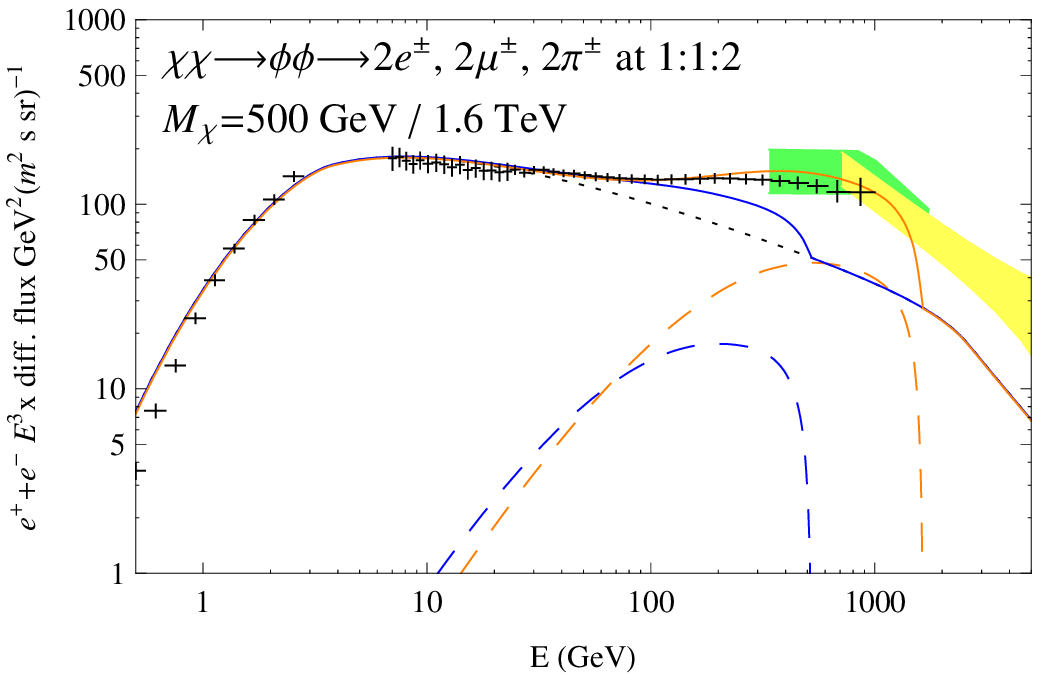}
\end{centering}
\caption{The same as in Figs.~1 and~2, but for a diffusion zone half-width of $L=2$ kpc. For annihilations to $2e^+ 2e^-$ and a mass of 400 GeV (1200 GeV), we have used a thermally averaged annihilation cross section of $\langle \sigma v \rangle = 1.2\times 10^{-25}$ cm$^3$/s ($1.2 \times 10^{-24}$ cm$^3$/s), with $\chi^{2}$/d.o.f. fit of 23(33) to the \textit{AMS} and 6.6(32) to the \textit{Fermi} data. For annihilations to $2\mu^+ 2\mu^-$ and a mass of 800 GeV (2.5 TeV), we have used a thermally averaged annihilation cross section of $5.6\times 10^{-24}$ cm$^3$/s ($5.1 \times 10^{-23}$ cm$^{3}$/s) resulting in $\chi^{2}$/d.o.f. fit of 13.1(30) to the \textit{AMS} and 9.5(7.3) to the \textit{Fermi} data. For annihilations to $2\pi^+ 2\pi^-$ and a mass of 1.0 TeV (3.0 TeV), we have used a thermally averaged annihilation cross section of $9.8\times 10^{-24}$ cm$^3$/s ($7.8 \times 10^{-24}$ cm$^3$/s) with a $\chi^{2}$/d.o.f. fit of 9.9(25) to the \textit{AMS} and 6.8(4.2) to the \textit{Fermi} data. For annihilations to a 1:1:2 ratio of $e^+ e^-$, $\mu^+ \mu^-$, and $\pi^+ \pi^-$ final states with a mass of 500 GeV (1.6 TeV), we have used a thermally averaged annihilation cross section of $2.7\times 10^{-24}$ cm$^3$/s ($2.3 \times 10^{-23}$ cm$^{3}$/s) resulting in a $\chi^{2}$/d.o.f. fit of 7.3(22) to the \textit{AMS} and 8.9(0.70) to the \textit{Fermi} data.}
\label{fig:KRA2kpc_DM}
\end{figure*}

\begin{figure*}
\begin{centering}
\includegraphics[width=3.30in,angle=0]{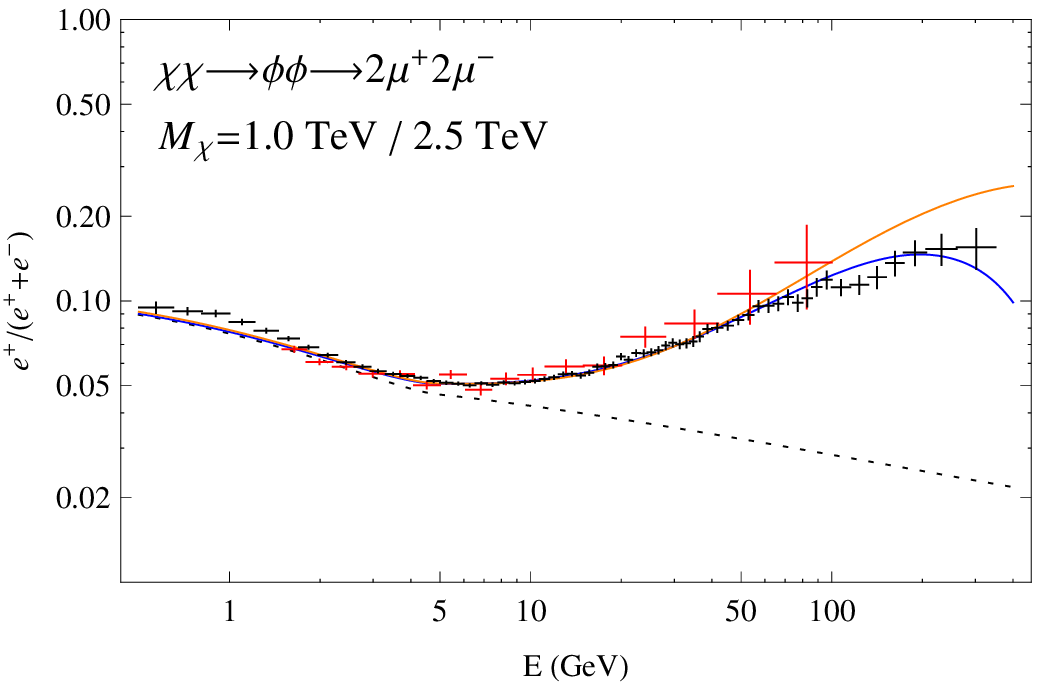}
\includegraphics[width=3.30in,angle=0]{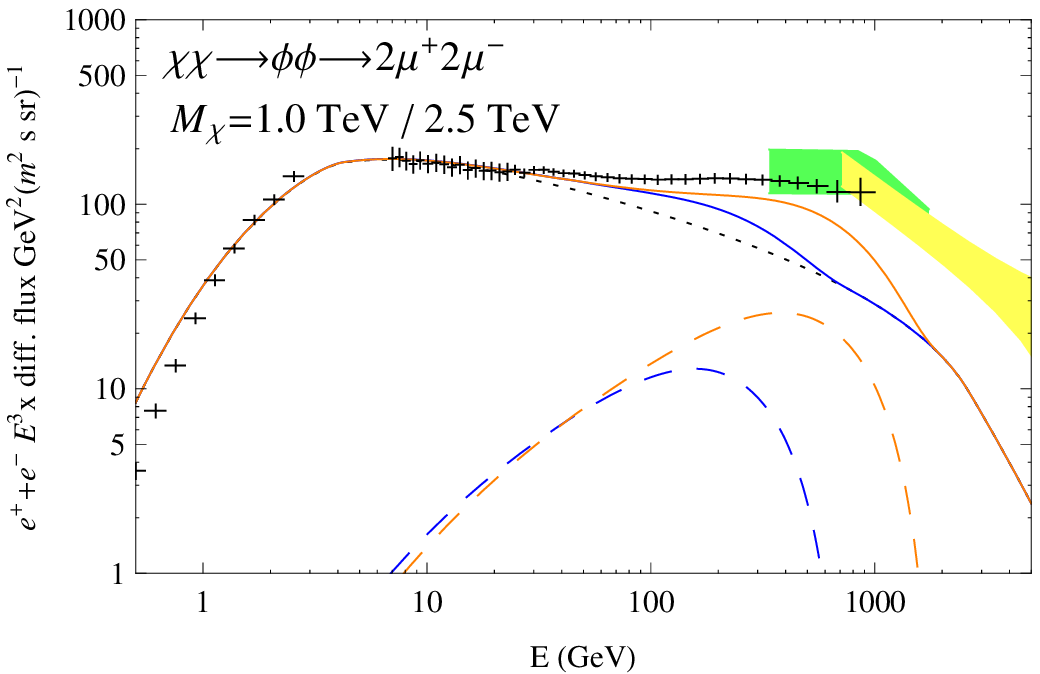} \\
\includegraphics[width=3.30in,angle=0]{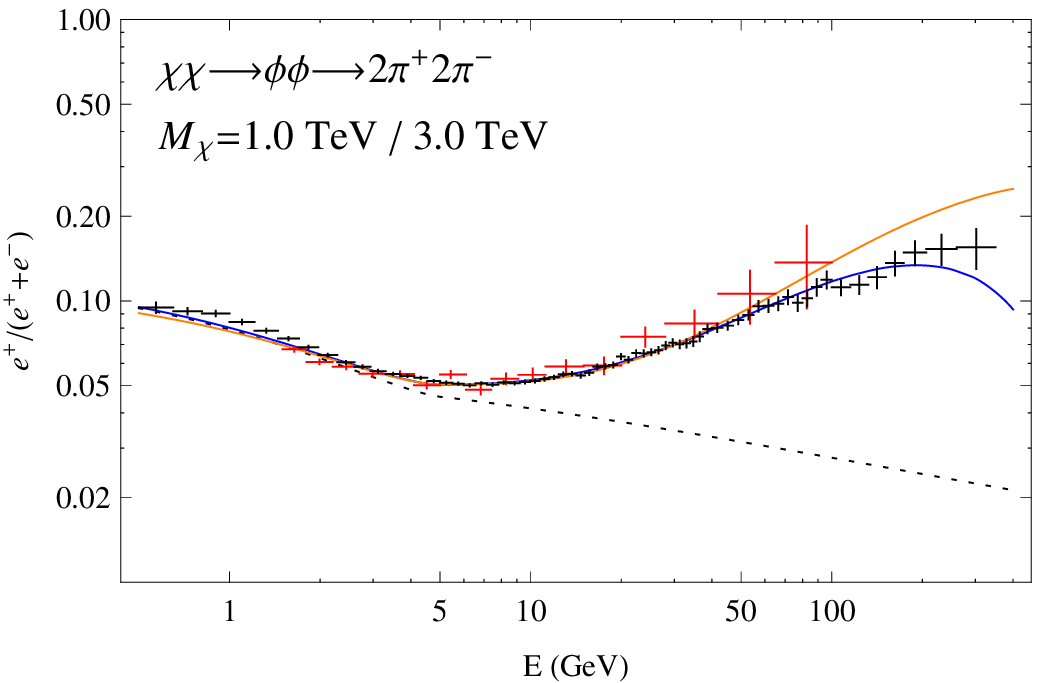}
\includegraphics[width=3.30in,angle=0]{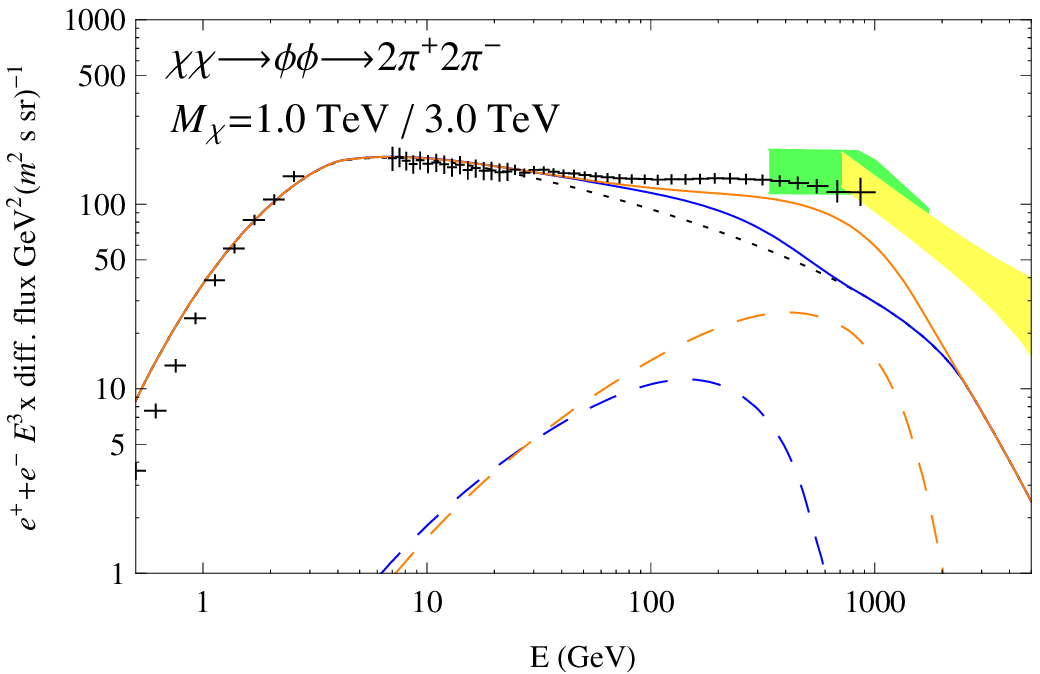} \\
\includegraphics[width=3.30in,angle=0]{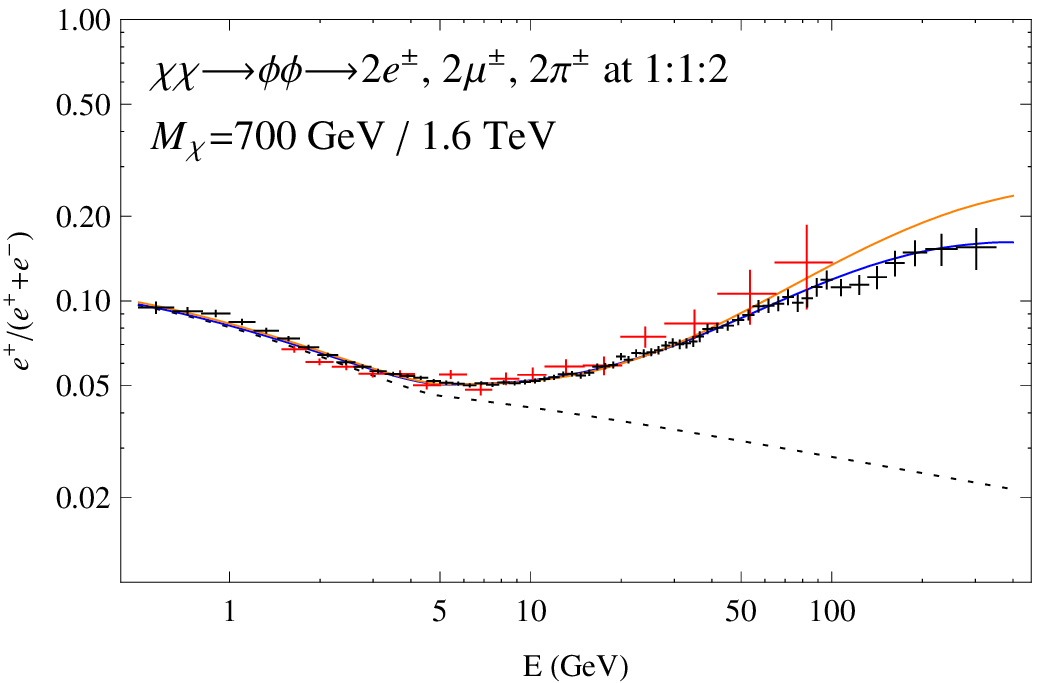}
\includegraphics[width=3.30in,angle=0]{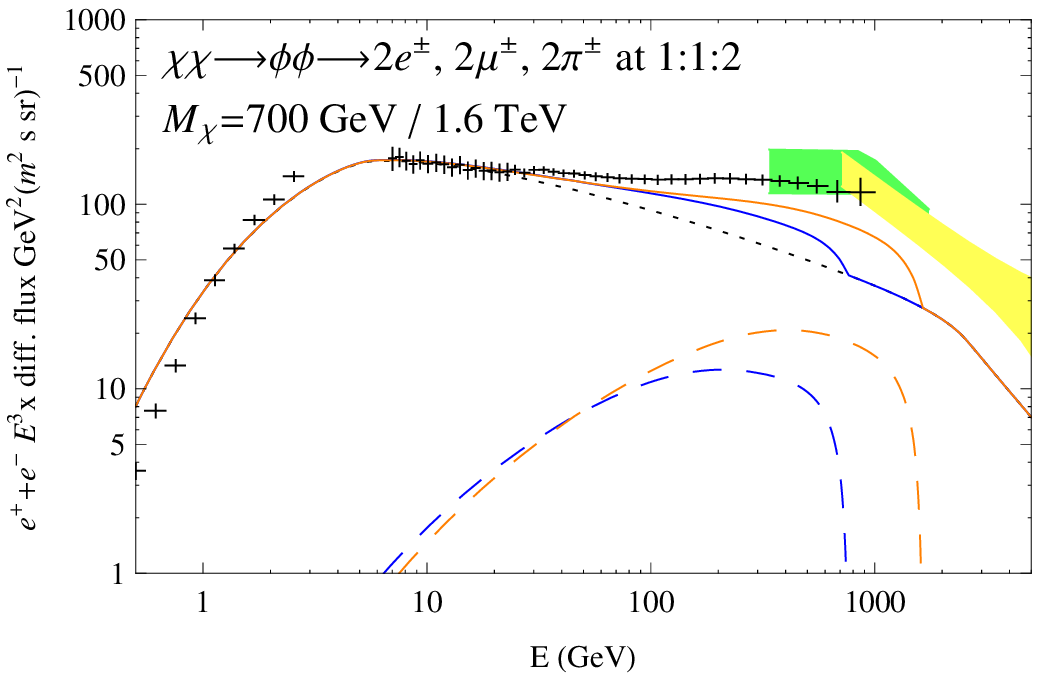}
\end{centering}
\caption{The same as in Figs.~1,~2 and~4, but for a diffusion zone half-width of $L=8$ kpc. For annihilations to $2\mu^+ 2\mu^-$ and a mass of 1.0 TeV (2.5 TeV), we have used a thermally averaged annihilation cross section of $2.9\times 10^{-24}$ cm$^3$/s ($1.5 \times 10^{-23}$ cm$^{3}$/s) providing a $\chi^{2}$/d.o.f. fit of 1.18(4.4) to the \textit{AMS} and 15.4(5.1) to the \textit{Fermi} data. For annihilations to $2\pi^+ 2\pi^-$ and a mass of 1.0 TeV (3.0 TeV), we have used a thermally averaged annihilation cross section of $3.5\times 10^{-24}$ cm$^3$/s ($2.3 \times 10^{-23}$ cm$^3$/s) giving a $\chi^{2}$/d.o.f. fit of 0.81(3.9) to the \textit{AMS} and 15.2(3.8) to the \textit{Fermi} data. For annihilations to a 1:1:2 ratio of $e^+ e^-$, $\mu^+ \mu^-$, and $\pi^+ \pi^-$ final states with a mass of 700 GeV (1.6 TeV), we have used a thermally averaged annihilation cross section of $1.6\times 10^{-24}$ cm$^3$/s ($6.5 \times 10^{-24}$ cm$^{3}$/s) with a $\chi^{2}$/d.o.f. fit of 0.83(3.0) to the \textit{AMS} and 13.4(7.6) to the \textit{Fermi} data.}\label{fig:KRA8kpc_DM}
\end{figure*}

By considering different values of the diffusion zone thickness, $L$, we can slightly alter the predicted shape of the positron fraction, potentially enabling some DM models to better accommodate the measurements of \textit{AMS}. In Fig.~\ref{fig:XDMmu_DiffZones}, we show for a representative DM model how changing this parameter impacts the shape of the positron fraction. Note that if we increase $L$ (to 8 kpc, for example), we can soften the slope of the rising positron fraction, which in some cases allows for a better fit to the data. In Figs.~\ref{fig:KRA2kpc_DM} and~\ref{fig:KRA8kpc_DM}, we show predictions for various DM scenarios using choices of $L=2$ and 8 kpc, respectively. As expected, the fits are worsened in the 2 kpc case, while in general improved for $L=8$ kpc, suggesting a preference for softer propagated $e^{\pm}$ spectra from DM annihilations. 

In none of the cases we have shown so far have we been able to find good agreement with the cosmic ray electron+positron spectrum as measured by \textit{Fermi}. To accommodate this measurement, we have to consider a cosmic ray electron spectrum which is not a simple power-law, but instead breaks to a harder spectral index at high energies. This is not surprising, as at energies above $\sim$100 GeV energy losses limit the number of sources that contribute to the electron cosmic ray spectrum; the cosmic ray electron spectrum at energies above a few hundred GeV is, therefore, likely to be dominated by only a small number of nearby sources. As a result, stochastic variations in the distribution of supernova remnants are expected to lead to local departures from the average cosmic ray spectrum found throughout the Milky Way (which may very well be a simple power-law)~\cite{Grasso:2009ma,Hooper:2009cs}. When the \textit{Fermi} electron+positron spectrum is taken in combination with the positron fraction as measured by \textit{AMS}, it is clear that more very high energy cosmic ray electrons are required than would be predicted using a simple power-law extrapolated from the low-energy spectrum.

\begin{figure*}
\begin{centering}
\includegraphics[width=3.30in,angle=0]{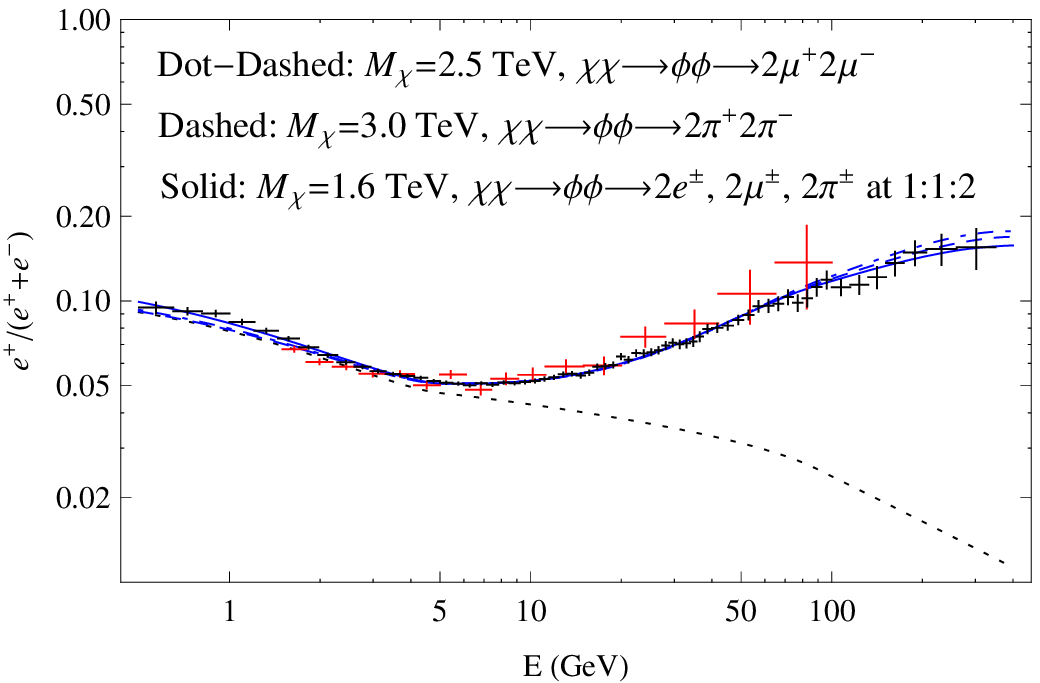}
\includegraphics[width=3.30in,angle=0]{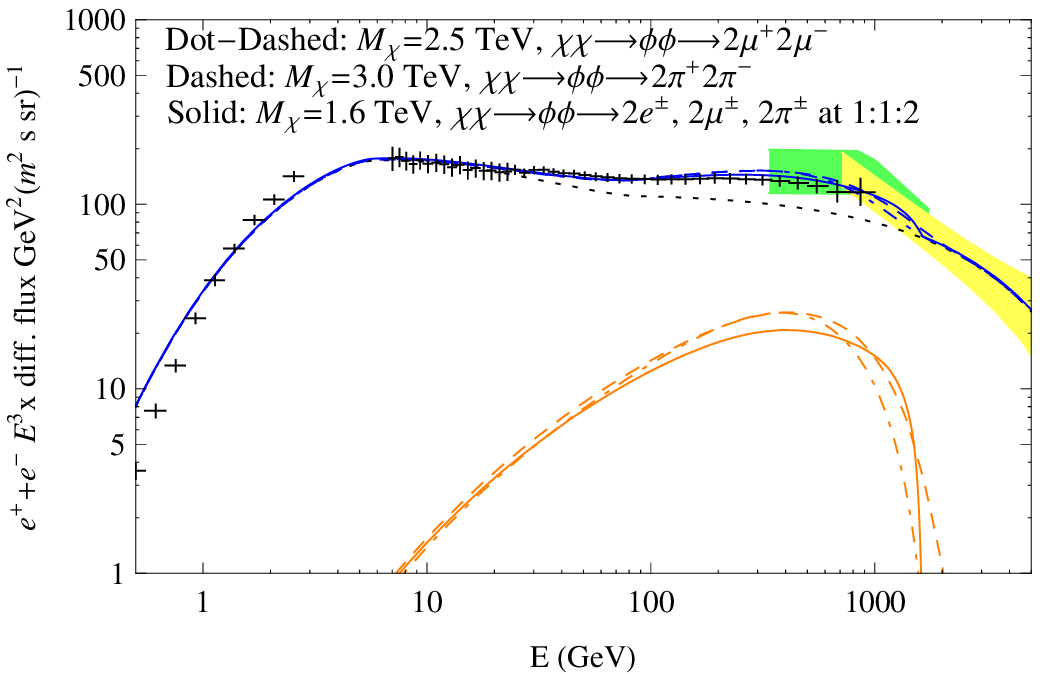} \\
\end{centering}
\caption{The same as in Figs.~1,~2,~4 and~5 but for a diffusion zone half-width of $L=8$ kpc, and for broken power-law spectrum of electrons injected from cosmic ray sources ($dN_{e^-}/dE_{e^-}\propto E_e^{-2.65}$ below 85 GeV and $dN_{e^-}/dE_{e^-}\propto E_e^{-2.3}$ above 85 GeV). The cross sections are the same as given in the caption of Fig.~5. With this cosmic ray background, we show the dark matter models compared to the measurements of the cosmic ray positron fraction and the overall leptonic spectrum. Even with the presence of a break, there is a preference towards models with softer injection $e^{\pm}$ spectra; with the 1.6 TeV to $e^{\pm}, \mu^{\pm}, \pi^{\pm}$ case providing the best $\chi^2 /d.o.f.$ fit to the \textit{AMS} (\textit{Fermi}) lepton data of 0.82(0.51). The 2.5 TeV to 2$\mu^{+}$ 2$\mu^{-}$, gives a $\chi^2 /d.o.f.$ fit of 1.32(1.07) and the 3.0 TeV to 2$\pi^{+}$ 2$\pi^{-}$ a fit of 1.00(1.03). We remind that in the \textit{Fermi} error-bars we do not include an overall shift from the energy resolution uncertainty.}
\label{fig:DMbreak}
\end{figure*}

In Fig.~\ref{fig:DMbreak}, we show results using a broken power-law for the spectrum of electrons injected from cosmic ray sources, for the three DM models most capable of accommodating the observed positron fraction. Between 4 and 100 GeV, we take the injected electron spectrum to be $dN_e/dE_e \propto  E_e^{-2.65}$ which provides a good fit to the observed low energy spectrum. Above 100 GeV, we harden this slope from -2.65 to -2.3.\footnote{Note that instead of a hardening of the injected electron spectrum above $\sim$100 GeV, we could have instead considered a mild overdensity of local sources. As electrons from local sources experience less energy loss than those from more distant sources, the contribution from local sources will exhibit a harder spectrum, which could dominate the observed electron spectrum at high energies even if the shape of the electron spectrum injected from these sources is the same as that from the average source. 
We also note that the slope of -2.65 bellow $\sim$ 100 GeV for the primary component, is chosen to provide a good fit to the \textit{Fermi} data, and is generally considered to be steeper than what is anticipated from 1st order Fermi accelerated electrons ($\lsim$ -2.0) \cite{DiBernardo:2010is}.}  With this spectral break, these three DM models now each appear to be capable of self-consistently accounting for both the measured positron fraction and the overall leptonic cosmic ray spectrum. Their $\chi^{2} /$d.o.f. fits are 1.32 / 1.00 / 0.82 to the \textit{AMS} data and 1.07 / 1.03 / 0.51 to the \textit{Fermi} data for the XDM to 2$\mu^{\pm}$, XDM to 2$\pi^{\pm}$ and XDM to 2$e^{\pm}$, 2$\mu^{\pm}$, 2$\pi^{\pm}$ at 1:1:2 relative branching ratios respectively. 

\section{Pulsars}
\label{sec:Pulsars}

Pulsars are rapidly spinning neutron stars which steadily convert their rotational kinetic energy into radio emission, $\gamma$-rays, and cosmic rays, likely including energetic electron-positron pairs. When initially formed, typical pulsars exhibit rotational periods on the order of tens or hundreds of milliseconds, and magnetic field strengths of $\sim$$10^{11}$-$10^{13}$ G. As a result of magnetic-dipole braking, the period of the pulsar's rotation slows down at a rate given by $\dot{P} = 3.3\times 10^{-15} \,(B/10^{12}\, {\rm G})^2 \, (P/0.3 \,{\rm s})^{-1}$, corresponding to an energy loss rate of $\dot{E}=4 \pi^2 I \dot{P}/P^3 = 4.8\times 10^{33} {\rm erg/s} \, (B/10^{12}\, {\rm G})^2 \, (P/0.3 \,{\rm s})^{-4} \, (I/10^{45} \rm{g}\,\rm{cm}^2)$. Young pulsars spin-down quite rapidly, typically losing a majority of their rotational kinetic energy in on the order of only $\sim$$10^5$ years. In both polar gap and outer gap models, a significant fraction of this energy can go into the production and acceleration of electron-positron pairs~\cite{Ruderman:1975ju,1983ApJ...266..215A,Cheng:1986qt}.

The spectral shape of the electrons and positrons injected from pulsars is often parametrized as~\cite{2001A&A...368.1063Z}:
\begin{equation}
\frac{dN_e}{dE_e} \propto E^{-\alpha}_e \exp(-E_e/E_c).
\end{equation}
Although there is considerable uncertainty associated with these spectral parameters, $\alpha = 1.5-2.0$ and $E_c = 80-1000$ GeV cover the range typically found throughout the literature. We begin by calculating the contribution from the sum of all pulsars distributed throughout the Milky Way. To do this, we adopt the spatial distribution of the pulsar birth rate as described in Ref.~\cite{FaucherGiguere:2005ny}, normalized to an overall rate of 1 pulsar per century throughout the Galaxy, each with an average total energy of $10^{49}$ erg. 

\begin{figure*}
\begin{centering}
\includegraphics[width=3.30in,angle=0]{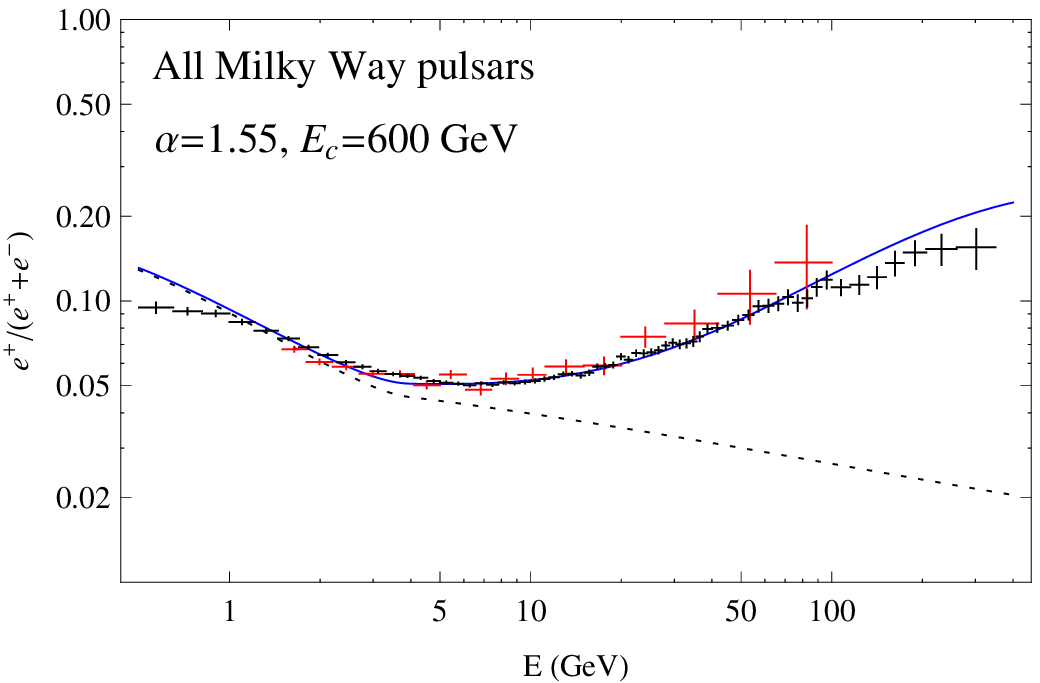}
\includegraphics[width=3.30in,angle=0]{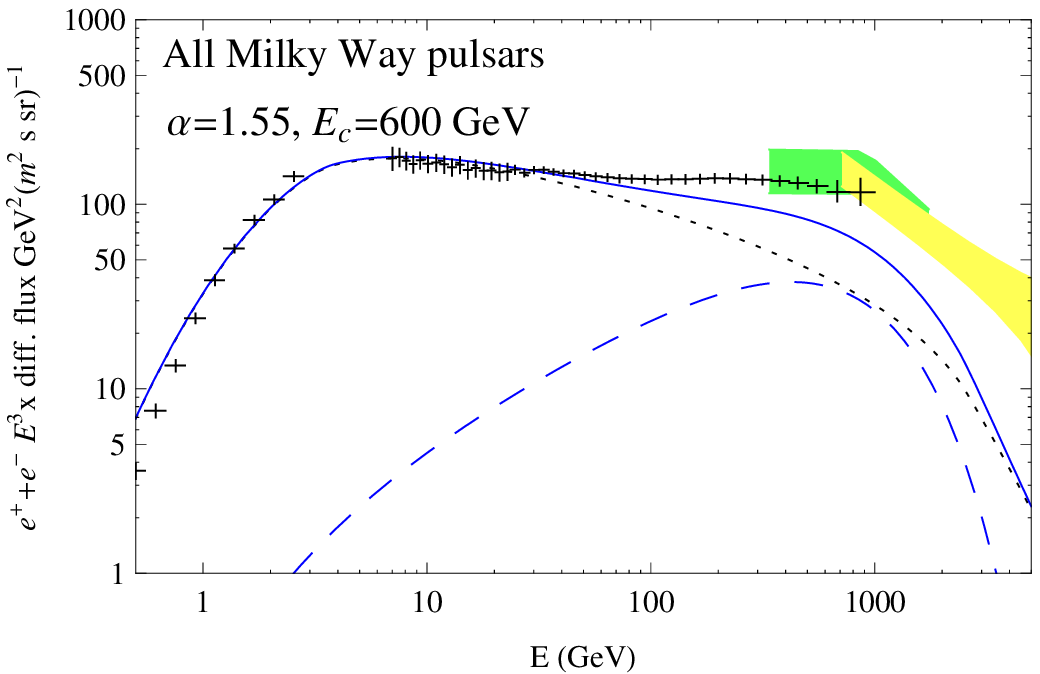}\\
\includegraphics[width=3.30in,angle=0]{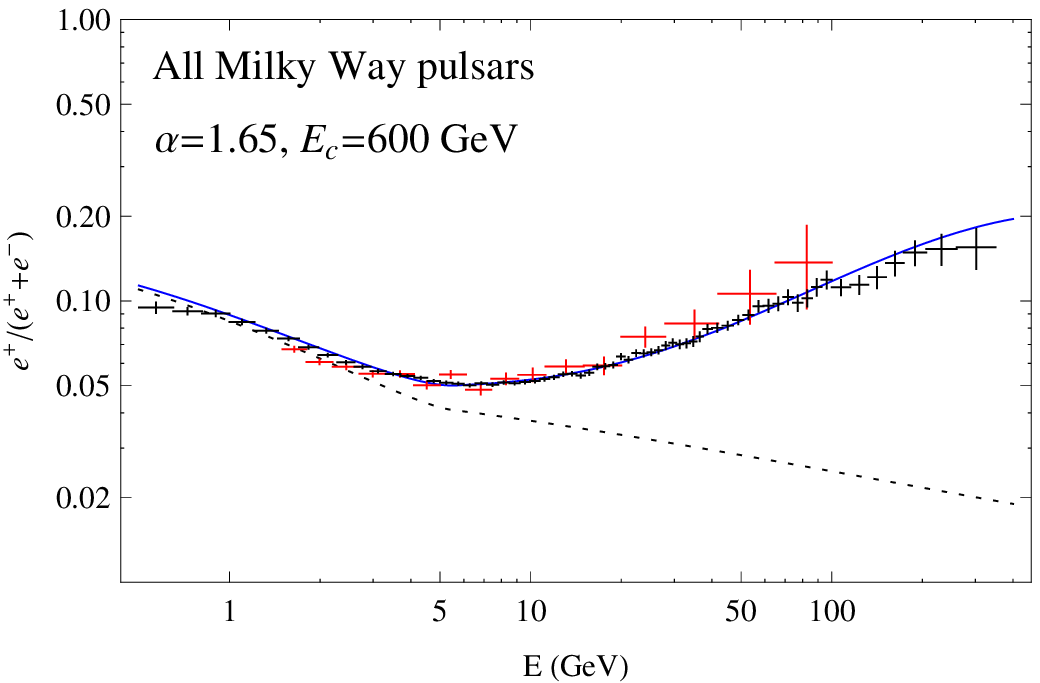}
\includegraphics[width=3.30in,angle=0]{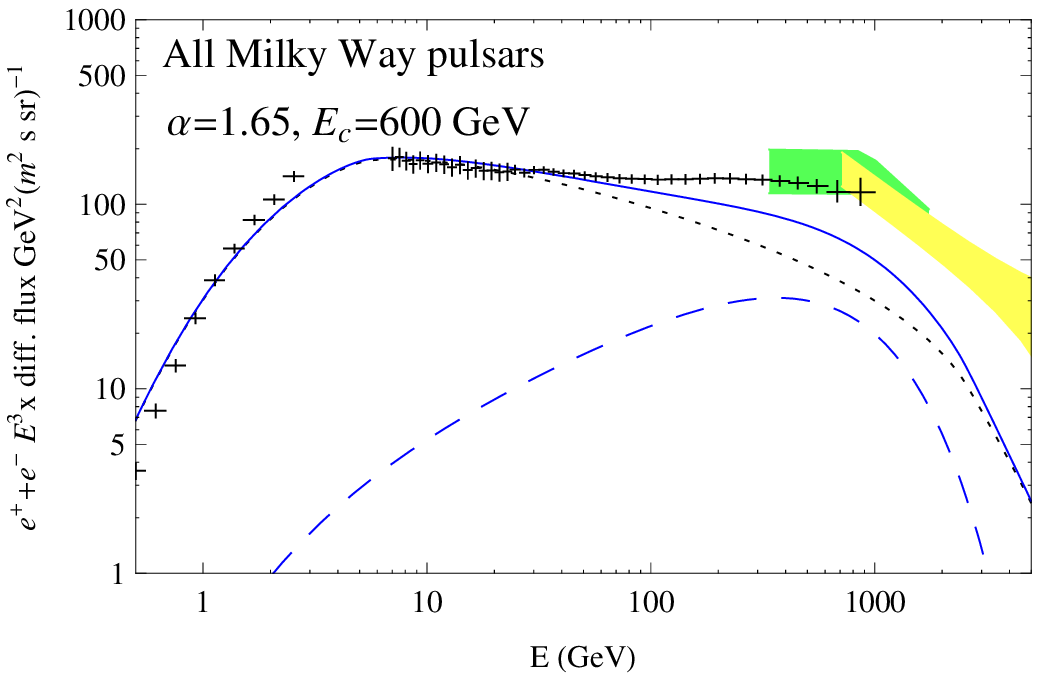}
\end{centering}
\caption{The predicted cosmic ray positron fraction (left) and electron+positron spectrum (right) from the sum of all pulsars throughout the Milky Way, for an injected spectrum of $dN_{e^{\pm}}/dE_{e^{\pm}} \propto E_{e^{\pm}}^{-1.55} \exp(-E_{e^{\pm}}/600 \,{\rm GeV})$ and a diffusion zone half-width of $L=4$ kpc (top), and for an injected spectrum of $dN_{e^{\pm}}/dE_{e^{\pm}} \propto E_{e^{\pm}}^{-1.65} \exp(-E_{e^{\pm}}/600 \,{\rm GeV})$ and a diffusion zone half-width of $L=8$ kpc (bottom).
 For normalization, we have assumed that 16\% of the pulsars' total energy goes into high energy electron-positron pairs. The error bars shown represent the positron fraction as measured by \textit{AMS} (black, left) and \textit{PAMELA} (red, left), and the electron+positron spectrum as measured by \textit{Fermi} and \textit{AMS-01} (black, right). In each case, we have adopted a propagation model that provides a good fit to the various secondary-to-primary ratios as described in the text and a $\chi^{2}$/d.o.f. fit to the \textit{AMS} data of 1.69(top), 1.11(botom) and 6.9(top) 8.7(bottom) to the \textit{Fermi} data. The expected backgrounds are shown as black dotted lines.}
\label{pulsarfig}
\end{figure*}

\begin{figure*}
\begin{centering}
\includegraphics[width=3.30in,angle=0]{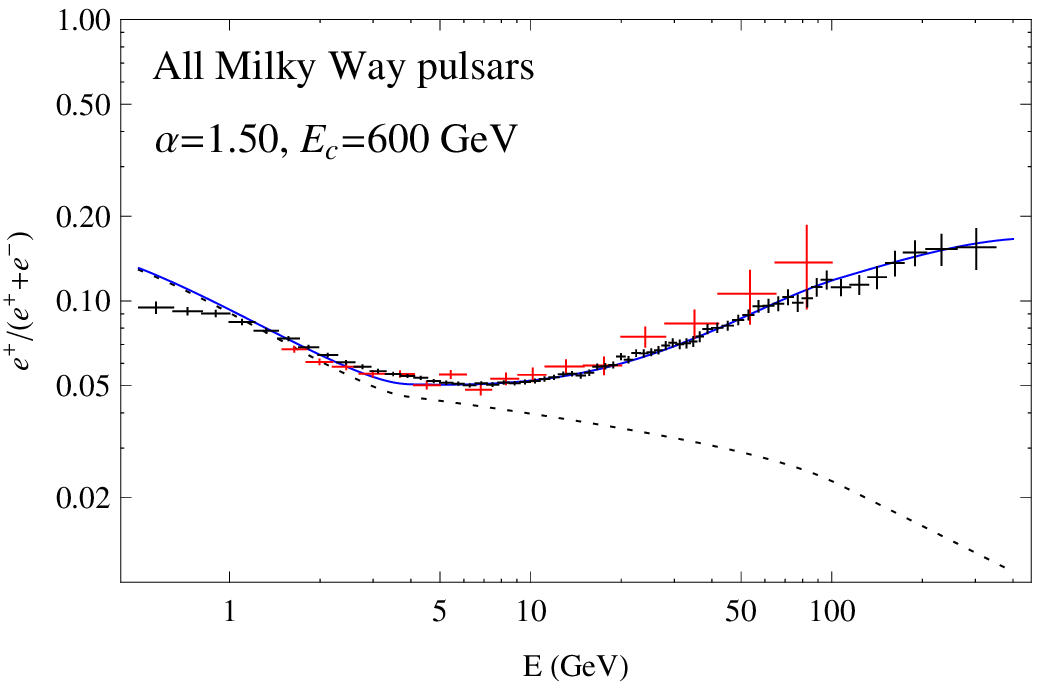}
\includegraphics[width=3.30in,angle=0]{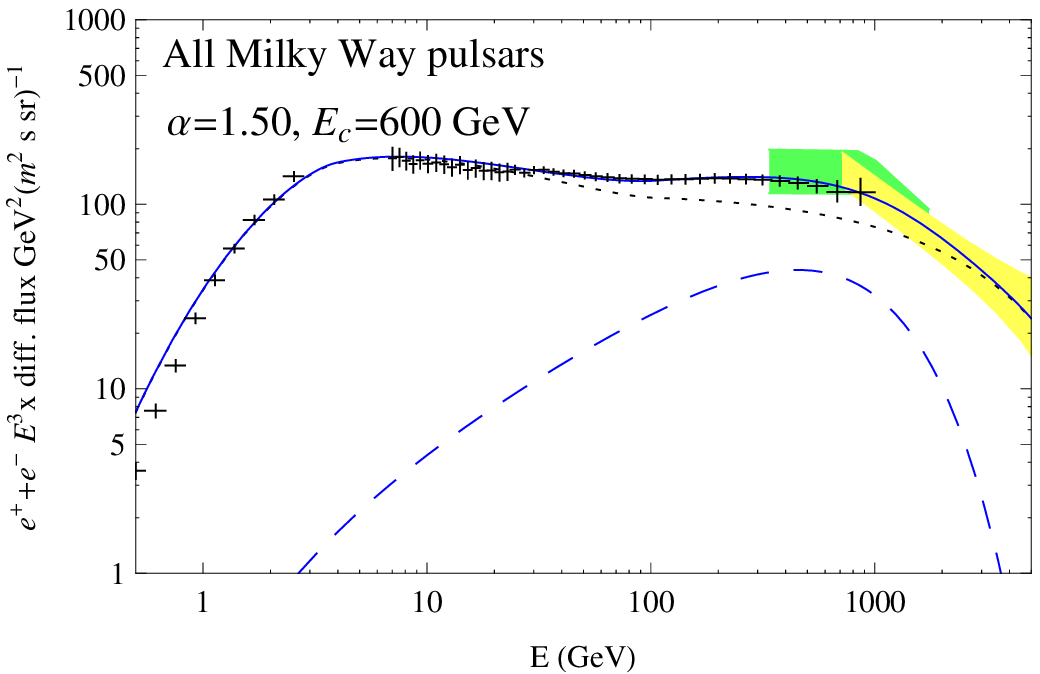}\\
\includegraphics[width=3.30in,angle=0]{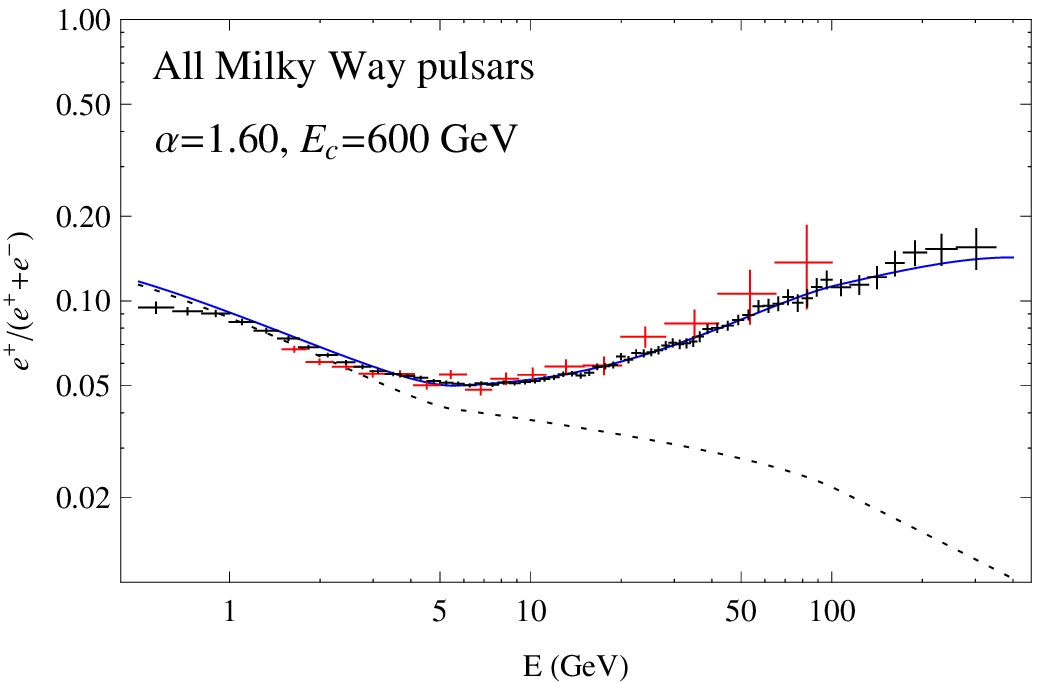}
\includegraphics[width=3.30in,angle=0]{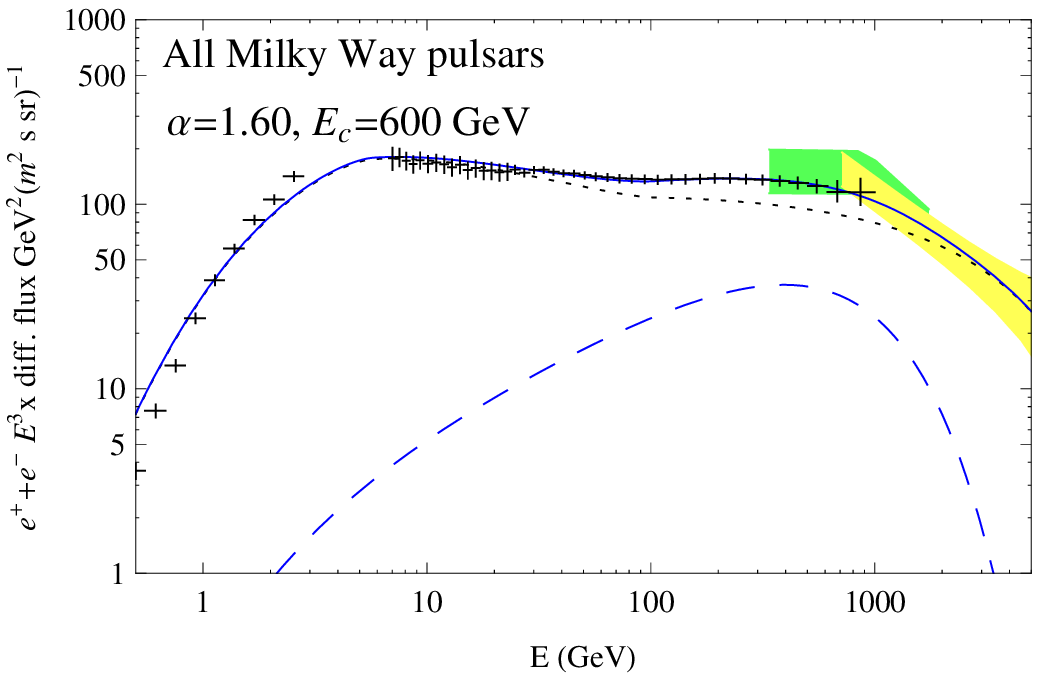}
\end{centering}
\caption{The same as in Fig.~\ref{pulsarfig}, but for a broken power-law spectrum of electrons injected from cosmic ray sources ($dN_{e^-}/dE_{e^-}\propto E_e^{-2.65}$ below 100 GeV and $dN_{e^-}/dE_{e^-}\propto E_e^{-2.3}$ above 100 GeV), and for slightly different pulsar spectral indices ($\alpha=$1.6 and 1.5 in the upper and lower frames, respectively). With this cosmic ray background, the pulsar models shown can simultaneously accommodate the measurements of the cosmic ray positron fraction and the overall leptonic spectrum giving a $\chi^{2}$/d.o.f. fit to the \textit{AMS} data of 0.85(top), 0.88(botom) and 0.37(top) 0.37(bottom) to the \textit{Fermi} data. By comparing to the results of Fig..~\ref{pulsarfig} the presence of a break at $\simeq$ 100 GeV is preferred from both individual data sets and from their combination.}
\label{pulsarfigbreak}
\end{figure*}

Following the procedure followed in the previous section, in Figs.~\ref{pulsarfig} and~\ref{pulsarfigbreak} we show results for this distribution of pulsars. In Fig.~\ref{pulsarfig}, we adopt a simple power-law of index -2.65 for the injected spectrum from (non-pulsar) cosmic ray sources, while in Fig.~\ref{pulsarfigbreak} we break this spectrum to -2.3 above 100 GeV. In each figure, we show results for two different choices of the pulsar spectral index, $\alpha$, and the diffusion zone half-width, $L$. In each case, we fix $E_{c}=600$ GeV and normalize the positron and electron contribution from pulsars by assuming that 16\% of the pulsars' total energy goes into high energy pairs. From these results (especially those shown in Fig.~\ref{pulsarfigbreak}), we conclude that for very reasonable choices of parameters, pulsars can provide a viable explanation for the observed cosmic ray positron fraction.

In addition to the integrated contribution from all pulsars throughout the Milky Way, there are two young and nearby pulsars which could each individually contribute significantly to the cosmic ray positron spectrum. The Geminga pulsar is 3.7$\times 10^5$ years old, 157 parsecs from the solar system, and pulsates with a period of 230 ms. The pulsar B0656+14 (possibly associated with the Monogem supernova remnant) is considerably younger (1.1$\times 10^{5}$ years), although somewhat more distant (290 parsecs), and more slowly rotating ($P=390$ ms). These parameters, combined with their measurements of $\dot{P}$, imply that Geminga and B0656+14 have each lost approximately $3\times 10^{49}$ erg and $1\times 10^{49}$ erg of rotational energy since their births, respectively. If 4-5\% of this energy went into the production and acceleration of energetic $e^+ e^-$ pairs, then these pulsars could be responsible for the observed rise in the cosmic ray positron fraction~\cite{Hooper:2008kg,Yuksel:2008rf}. If we combine these two sources with the somewhat smaller contribution expected from the sum of all more distant pulsars~\cite{Hooper:2008kg}, we estimate that if 3-4\% of the total energy from pulsars goes into energetic pairs, this would be sufficient to account for the observed positrons.

\section{Summary and Discussion}
\label{sec:Conclusions}

In this paper, we have revisited both annihilating dark matter and pulsars as possible sources of the rising cosmic ray positron fraction. Using the newly published, high precision data from \textit{AMS}, we have considered a wide range of dark matter models and cosmic ray propagation models. We find that models in which the dark matter annihilates directly to leptons ($e^+ e^-$ or $\mu^+ \mu^-$) are no longer capable of producing the observed rise in the positron fraction. Models in which the dark matter annihilates into light intermediate states which then decay into combinations of muons and charged pions, however, can accommodate the new data (see Fig.~\ref{fig:DMbreak}). In those dark matter models still capable of generating the observed positron excess, the dark matter's mass and annihilation cross section fall in the range of $\sim$1.5-3 TeV and $\langle \sigma v \rangle \sim (6-23)\times 10^{-24}$ cm$^3$/s.

We have also considered pulsars as a possible source of the observed positrons. In particular, we find that for reasonable choices of spectral parameters and spatial distributions, the sum of all pulsars in the Milky Way could account for the observed positrons (see Fig.~\ref{pulsarfigbreak}) if, on average, 10-20\% of their total energy goes into the production and acceleration of electron-positron pairs (assuming a birth rate of one per century throughout the Galaxy, each with an average total energy of $10^{49}$). It may also be the case that a small number of nearby and young pulsars (most notably Geminga and B0656+14) could dominate the local cosmic ray positron flux at energies above several tens of GeV. Taking into account these two exceptional sources, we estimate that if 3-4\% of the total energy from pulsars goes into energetic pairs, these objects could be responsible for the observed positron fraction.

Currently, we cannot yet discriminate between dark matter and pulsars as the source of the observed positron excess. We are hopeful, however, that future data from \textit{AMS} may change this situation. In addition to continuing to improve the precision of their measurement of the positron fraction and extending this measurement to higher energies, \textit{AMS} will also measure with unprecedented precision a number of secondary-to-primary ratios of cosmic ray nuclei species, which can be used to constrain many aspects of the underlying cosmic rays propagation model. Of particular importance is the $^{10}$Be/$^9$Be ratio, for which existing measurements are limited to energies below 2 GeV (kinetic energy per nucleon), and with large errors (for a compilation of such measurements, see Tables I and II of Ref.~\cite{Simet:2009ne}). In contrast, \textit{AMS} is expected to measure this ratio with much greater precision, and up to energies of $\sim$10 GeV. This information will enable us to break the longstanding degeneracy between the diffusion coefficient and the boundary conditions of the diffusion zone~\cite{Pato:2010ih}. If these measurements ultimately favor propagation models with a somewhat narrow diffusion zone ($L \lsim 4$ kpc), it would be very difficult to explain the observed positron fraction with any of the dark matter models we have considered in this paper, while pulsar models could still provide a viable source for the positrons. Thus, in at least some plausible scenarios, future data from \textit{AMS} could allow us to rule out competing hypotheses and thereby finally reach a conclusion as to the origin of the surprisingly large flux of energetic cosmic ray positrons.

\vskip 0.2 in
\section*{Acknowledgments}  
While completing this project, a paper addressing similar issues appeared~\cite{Yuan:2013eja}. This work has been supported by the US Department of Energy. 
\vskip 0.05in

\bibliography{AMS02positrons}
\bibliographystyle{apsrev}

\end{document}